\documentclass[10pt,journal,compsoc]{IEEEtran}
\ifCLASSOPTIONcompsoc
  \usepackage[nocompress]{cite}
\else
  \usepackage{cite}
\fi
\ifCLASSINFOpdf
\else
\fi
\usepackage{algorithm}
\usepackage[noend]{algpseudocode}
\usepackage{color,soul}
\usepackage{tikz}
\usetikzlibrary{shapes}
\usepackage{hyperref}
\usepackage{amsmath}
\usepackage{amssymb}

\usepackage{tabu}
\usepackage{threeparttable}
\usepackage{threeparttablex}
\usepackage[normalem]{ulem}
\usepackage{makecell}
\usepackage{xcolor}

\usepackage{enumitem}

\newcommand{\SYSTEMNAME}{{\texttt{GEViTRec}}}

\newcommand{\spoke}{\raisebox{0.5pt}{\tikz{\node[draw,scale=0.6,circle,fill=none](){};}}}
\newcommand{\hub}{\raisebox{0.0pt}{\tikz{\node[draw,scale=0.8,regular polygon, regular polygon sides=4,fill=none](){};}}}

\newcommand{\rev}[1]{#1}
\newcommand{\ac}[1]{}
\newcommand{\tm}[1]{#1}

\begin{document}
\title{GEViTRec: Data Reconnaissance Through Recommendation Using a Domain-Specific Prevalence Visualization Design Space}

\author{Anamaria Crisan,
        Shannah E. Fisher,
        Jennifer L. Gardy,
        and~Tamara~Munzner,~\IEEEmembership{Senior~Member,~IEEE}% <-this % stops a space
\IEEEcompsocitemizethanks{
\IEEEcompsocthanksitem A. Crisan is with Tableau Research. Email: acrisan@tableau.com.\protect\\
\IEEEcompsocthanksitem S. Fisher and T. Munzner are with the University of British Columbia. Email: shannahelizabeth@gmail.com, tmm@cs.ubc.ca.\protect\\
\IEEEcompsocthanksitem J. Gardy is with the Gates Foundation. Email: jennifer.gardy@gatesfoundation.org}% <-this % stops an unwanted space
\thanks{Manuscript received September 15, 2020; revised TBD.}}

\markboth{TVCG Submission}%
{Crisan \MakeLowercase{\textit{et al.}}: GEViTRec}



\IEEEtitleabstractindextext{%
\begin{abstract}
\rev{Genomic Epidemiology (genEpi) is a branch of public health that uses many different data types including tabular, network, genomic, and geographic, to identify and contain outbreaks of deadly diseases. Due to the volume and variety of data, it is challenging for genEpi domain experts to conduct data reconnaissance; that is, have an overview of the data they have and make assessments toward its quality, completeness, and suitability. We present an algorithm for data reconnaissance through automatic visualization recommendation, GEViTRec. Our approach handles a broad variety of dataset types and automatically generates coordinated combinations of charts, in contrast to existing systems that primarily focus on singleton visual encodings of tabular datasets. We automatically detect linkages across multiple input datasets by analyzing non-numeric attribute fields, creating an entity graph within which we analyze and rank paths. For each high-ranking path, we specify chart combinations with spatial and color alignments between shared fields, using a gradual binding approach to transform initial partial specifications of singleton charts to complete specifications that are aligned and oriented consistently. A novel aspect of our approach is its combination of domain-agnostic elements with domain-specific information that is captured through a domain-specific visualization prevalence design space. Our implementation is applied to both synthetic data and real data from an Ebola outbreak. We compare GEViTRec's output to what previous visualization recommendation systems would generate, and to manually crafted visualizations used by  practitioners. We conducted formative evaluations with ten genEpi experts to assess the relevance and interpretability of our results. \\
\newline
\textbf{Code, Data, and Study Materials Availability:} https://github.com/amcrisan/GEVitRec}
\end{abstract}

}

\maketitle

\IEEEdisplaynontitleabstractindextext

% \fi
\IEEEpeerreviewmaketitle

\IEEEraisesectionheading{\section{Introduction}\label{sec:introduction}}

\IEEEPARstart{D}{ata} \tm{reconnaissance was recently defined as the process of exploring a group of datasets that are not yet understood by a specific person, and an iterative four-stage process of acquire, view, assess, pursue was proposed as a conceptual framework to reason about the needs of the \textbf{data recon} process~\cite{DRTW_crisan_2019}. Although the visualization literature includes many systems designed for investigative exploration, where a specific existing dataset is transformed and analyzed in depth, the goal of very quickly understanding potential linkages between unfamiliar datasets remains difficult to achieve with existing systems.} 

\tm{We posit that visualization recommendation systems have great promise for operationalizing the goal of data recon through a concrete algorithm that quickly and automatically computes reasonable visual encodings with minimal input from a user. A recommender system could speed up the \textit{view} stage of the data recon process, in contrast to any kind of design or selection process for visual encoding that involves human judgement. The aim of data recon is to quickly assess whether the current collection of datasets is suitable for some intended analysis question, and if not to pursue additional data. After the recon process concludes when an appropriate collection of datasets has been acquired, a more lengthly analysis process with traditional investigative exploration visualization tools could be launched, where human judgement from tool creators or tool users is appropriate to incorporate into visual encoding design choices.} 

\tm{Prior work takes a one-size-fits-all approach based solely on domain-agnostic perceptual effectiveness rankings, from the foundational APT system~\cite{Mackinlay1986} to Tableau's ShowMe~\cite{mackinlay_show_2007} to recent activity including Voyager~\cite{Wongsuphasawat2017} and Draco~\cite{Moritz2019}. Such systems can help users rapidly understand their datasets by automatically suggesting suitable encodings based on data characteristics and the efficacy of different chart types. However, these systems primarily focus on encoding a single dataset of tabular data type. They fall short for two key requirements for the data recon use case: the ability to both find and show linkages between multiple datasets, and the ability to handle a broad variety of data types as input and a broad variety of chart types as output.}

\tm{A novel feature of our approach is that we tailor the recommendation for a specific domain, to tame the combinatorial explosion of possibilities that arise from a broad variety of input data types and output chart types. We leverage the knowledge encapsulated in a recent \textit{domain-specific visualization prevalence design space (VPDS)}, where visualization strategies used by experts in a particular domain are both characterized and enumerated~\cite{gevit2018}. We use this domain-specific information in a few targeted stages of our recommender algorithm, in conjunction with many domain-agnostic decision procedures. The GEViT~\cite{gevit2018} VPDS proposed in previous work arises from \textit{genomic epidemiology (genEpi)}, a very appropriate exemplar domain for data recon because it entails a diverse and heterogeneous set of input data types and output chart types, and genEpi analysts often grapple with unfamiliar collections of datasets in situations where access is tightly controlled due to sensitive personally identifiable information.}

\tm{The main contribution of our work is the design and implementation of \SYSTEMNAME, an end-to-end recommender algorithm to support the data recon process that computes potential linkages between datasets and automatically generates coordinated combinations of charts to illustrate them. The charts are coordinated through spatial layout and color alignment in the visual encodings of the individual charts, which are then arranged into a visually coherent combination. The resulting specification is rendered into a single static image of the coordinated combination that allows the linkages to be immediately perceived, without requiring any kind of time-consuming interactive exploration~\cite{Lam:2008:AFI}. Many stages of our proposed recommender algorithm feature domain-agnostic computations while some stages leverage domain-specific prevalence information. We validate our approach by comparing our results to existing genEpi visualization dashboards created by careful human curation. We also conducted an evaluative interview study with ten genEpi experts to verify the interpretability and utility of \SYSTEMNAME's results.}

\section{Domain-Specific Prevalence for Genomic Epidemiology}\label{sec:domain}

\rev{Visualization has historically been a prominent component of epidemiological research and practice since John Snow's infamous 1854 cholera map. More recently there has been an influx of new, heterogeneous, and multidimensional sources of data, including whole genome sequencing data and pulsed-field gel electrophoresis for DNA fingerprinting, that has birthed a new field: genomic epidemiology (genEpi)~\cite{Gardy2018}. Genomic data can be difficult to integrate with other data sources, including tabular data from electronic health records, network data from contact tracing, and spatial data~\cite{beliv2016,peerJ2018}.} \tm{The volume and variety of data at play in the genEpi domain make it an excellent exemplar for the utility of a domain-specific visualization prevalence design space in a proof-of-concept system. Moreover, there is a clear need for data recon in genEpi analysis contexts. The sensitivity of clinical health data leads to stringent access controls, so multiple rounds of acquire, view, assess, and pursue are often required to gather the requisite collection of datasets to answer any specific analysis question. Also, epidemiological analysis frequently takes place in an urgent context, particularly in pandemic times.}

\subsection{Genomic Epidemiology Visualization Typology}\label{sec:background}
%\ac{Formally, this was background and earlier. I went back to our original instincts from the VIS paper and put these implementation details....as implementation. I left then out of the technique description}.
%We briefly describe prior research findings that are necessary for understanding the design decisions of our technique.
A prior study characterized and enumerated the domain-specific data visualization strategies used by genEpi experts~\cite{gevit2018}. It proposed a \underline{G}enomic \underline{E}pidemiology \underline{Vi}sualization \underline{T}ypology (GEViT) that broke down how visualizations were constructed through chart types, enhancements, and combinations. The typology encompasses 25 unique chart types grouped within 8 categories (common statistical charts, color, relational, temporal, spatial, tree, genomic, and other), four types of chart combinations (spatially aligned, color aligned, small multiples, and unaligned), and two primary mechanisms of enhancements (adding or re-encoding marks). GEViT was developed using a corpus of approximately 18,000 research articles pertaining to genomic epidemiology that was representatively sampled to yield a set of 800 figures that informed the typology generation. Text mining techniques were applied to the titles and abstracts of all articles to derive a notion of the creation context for the sampled set of figures, so that connections between different data types and domain-specific attributes to data visualizations in the form of individual charts and chart combinations could be harvested. Most importantly, the representative sampling strategy also enabled the enumeration of these different visualization strategies, providing a quantitative measure of their relevance and importance to genEpi.
\tm{This information exactly constitutes a domain-specific visualization prevalence design space (VPDS), although that specific term is introduced here and was not used that work.}

\tm{We leverage the genEpi VPDS from the GEViT study in our proposed recommendation algorithm. The specific implementation of the GEViT design space draws visualizations from research articles, which are an instance of expert-to-expert communication. We posit that such a design space is suitable for generating visualization recommendations where an algorithm is attempting to communicate the data to experts to facilitate data reconnaissance. One very valuable property of the GEViT approach is that it captures visualizations made with many different systems, crucially including manual post-processing adjustments made by users with image creation and processing tools. Design spaces derived from analyzing visualizations made with only a single system, such as Tableau Public~\cite{Morton:2014} or ManyEyes~\cite{manyeyes} would not capture such a broad diversity of visualization strategies. Another important property is that the GEViT VPDS captures the year of use, providing longitudinal data about how and whether data visualization strategies might change over time.}

\subsection{Visualization Prevalence Design Spaces} \label{subsec:formalisms-designspace}

\tm{The fundamental idea behind our approach is to inject domain relevance into a visualization recommendation algorithm based on an analysis of the visualization design space in use in that domain. The term \textit{design space} is used broadly within the visualization research literature with many shades of meaning, and the GEViT approach that we leverage is one of many possible strategies for generating a visualization design space. Other examples of design spaces include SetVis~\cite{setvis2014}, TreeVis~\cite{schulz_treevis.net:_2011}, ManyEyes~\cite{manyeyes}, the Tableau Public Resources and analysis~\cite{Morton:2014}, and VizNet~\cite{Hu:2019:VTL:3290605.3300892}.}

Here, we define a \textbf{domain-specific design space} as a collection of \textbf{visual encodings $(V)$} that are produced by experts in some domain. We define a \textbf{domain-specific visualization prevalence design space $(D)$} as one that both captures the full scope (within reason) of visual encodings used by some definable set of experts and includes a quantitative estimate for prevalence of these visual encoding strategies in the domain. We use this quantitative prevalence information in our algorithm to define relevance scores for different visual encodings. Below, we will frequently abbreviate domain-specific visualization prevalence design space as simply a \textbf{design space}. 

\section{Related Work} \label{sec:related-work}
We survey related work on visualisation recommendation, on chart coordination and combinations, and on existing genEpi visualization tools.
%We consider related work from visualization recommendations and prior work on combinations as well as existing solutions in genEpi.

\subsection{Visualization Recommendation} \label{sec:related-recommender}
Automated recommendations of visual encodings are a way to help users explore datasets, and have been pursued by both the research and practitioner communities. A number of visualization recommender systems use rule based approaches that generate recommendations with the combination of a user's specifications and the types of data attributes (i.e. numeric, categorical, and nominal)~\cite{Mackinlay1986}. The ShowMe~\cite{mackinlay_show_2007} and Voyager~\cite{Wongsuphasawat2016,Wongsuphasawat2017} systems also rank visual encodings according to manually predetermined scores of graphical perception efficacy; highest ranked visualizations are prioritized and shown to the user, while lower ranked visualizations are accessible through interactions with the data visualization systems. The Draco system takes this ranking approach a step further by learning the efficacy scores automatically from the results of graphical perception experiments~\cite{Moritz2019}. Other recent systems, such as Data2Vis~\cite{data2vsi2018} and VizML~\cite{Hu:2019:VML:3290605.3300358}, also attempt to learn the associations between different types of data and visual encodings. Both of these systems analyze datasets themselves, not just attribute types, and discover the kinds of associations that users organically generate ``in the wild'' from publicly available resources. These learned associations are then used to automatically generate data visualizations. Another class of prior data visualization systems attempt to develop a semantic understanding of data and to use this knowledge in the recommendation processes. These systems include SemViz~\cite{semviz2008} and Cammarano's schema matching technique~\cite{Cammarano2007}; however, such techniques have not been expanded upon, likely due to the difficulties of  developing and maintaining ontologies. A final class of systems also attempts to infer user preferences via a collaborative filtering approach~\cite{Mutlu:2016}. The approaches taken by these different recommendation systems are not mutually exclusive and can be combined.

We take inspiration from these systems, but our technique extends beyond their capabilities in supporting a wider variety of data types and visual encodings, and in supporting the automatic creation of coordinated combinations of visual encodings. 

\subsection{Chart Coordination and Combination} \label{subsec:related-coordination}
Coordinated visual encodings are beneficial for highlighting shared attributes across a collection of datasets.  A great deal of previous work is devoted to interactive methods for linking views through shared information, particularly for juxtaposed views with linked highlighting~\cite{munzner_visualization_2015,Roberts2007}. However, coordinated multi-view displays that are static also play an important role and in fact represent a large portion of real-world use of data visualization -- yet their use and construction has been understudied by the research community~\cite{Sarikaya2019}. Several visualization authoring systems including Improvise~\cite{weaver_improvise}, Lyra~\cite{2014-lyra}, Data Illustrator~\cite{Liu:2018:}, and Charticulator~\cite{ren2019charticulator}, as well as visualization libraries like D3~\cite{bostock2010}, ggplot~\cite{ggplot}, and Vega~\cite{2017-vega-lite} (which underpins both Draco and Voyager), are able to support coordinated combinations of charts, including small multiples, aligned color palettes, and unaligned combinations. It is also possible to generate and coordinate single charts with commercial stand-alone applications such as Tableau, PowerBI, or Excel. However, analysts are particularly likely to use static rather than interactive charts~\cite{peerJ2018} when undertaking analyses that require programming in environments that integrate with sophisticated analysis methods, because the overhead of writing code for interactive view coordination tends to be high. To differentiate from interactive views, we deliberately use the term \textit{chart} to emphasize the static nature of these visualizations.

\tm{Despite the flexibility and popularity of view coordination, and prior work on grammars for facilitating coordinated combinations such as HiVE~\cite{Slingsby2009} and ATOM~\cite{Park2018}, constructing effective combinations remains a challenging proposition in practice~\cite{Qu2018Consistency}. No prior visualization recommendation systems attempts to do so. In contrast, \SYSTEMNAME\ treats coordinated combinations as first class citizens. To facilitate a consistent interface for coordinated and combined charts, our recommendation algorithm includes a specification syntax inspired by the GEViT~\cite{gevit2018} typology.} 

\subsection{Existing GenEpi Solutions}\label{subsec:related-domain}
GenEpi stakeholders can currently either develop their own solo or combined data visualizations using one of several charting libraries, or can use previously curated visualizations in publicly deployed dashboards and other interactive systems. We identified the charting libraries in common use:  TreeViewer~(stand-alone, \cite{huerta-cepas_ete_2016}), Baltic (python,~\cite{baltic}), ape~(R, \cite{apeRpck}), ggplot~(R, \cite{ggplot}), and  ggtree~(R, \cite{ggtree2017}). Since manually generating data visualizations is time consuming and a potential wasteful activity in the midst of serious epidemiological crises, people have also developed interactive systems and dashboards with manually pre-curated sets of visualizations that are updated as outbreaks evolve. We have identified  Nextstrain~\cite{hadfield_nextstrain:_2018}, Microreact ~\cite{Argimon2016} as two such widely used and state of the art systems, and compare the results of \SYSTEMNAME\ to them.

\section{Recommendation Algorithm}\label{sec:technique}
\begin{algorithm}[h!]
\caption{GEViTRec ($H$, $D$, $A$, $T$, $F_{users}$)}\label{alg:gevis-overview}
\begin{algorithmic}[1]
\Require Datasets $H$, Design Space $D$, Alignable Spatial Combinations (A), Chart Templates $T$, User Specified Fields $F_{users}$
\State (F,M) $\leftarrow$ \texttt{explodeFields}($H$)
\State E $\leftarrow$ \texttt{generateEntityGraph}($F$)
\State $R_p \leftarrow$ \texttt{rankPaths}($E,D$)
\State V $\leftarrow$ \texttt{generateVisSpecs}($R_p, M, F, F_{users}, A, T$)
\State \texttt{layoutAndRender}($V$)
\end{algorithmic}
\end{algorithm}

\begin{figure}[t!]
\centering % avoid the use of \begin{center}...\end{center} and use \centering instead (more compact)
\includegraphics[width=\columnwidth]{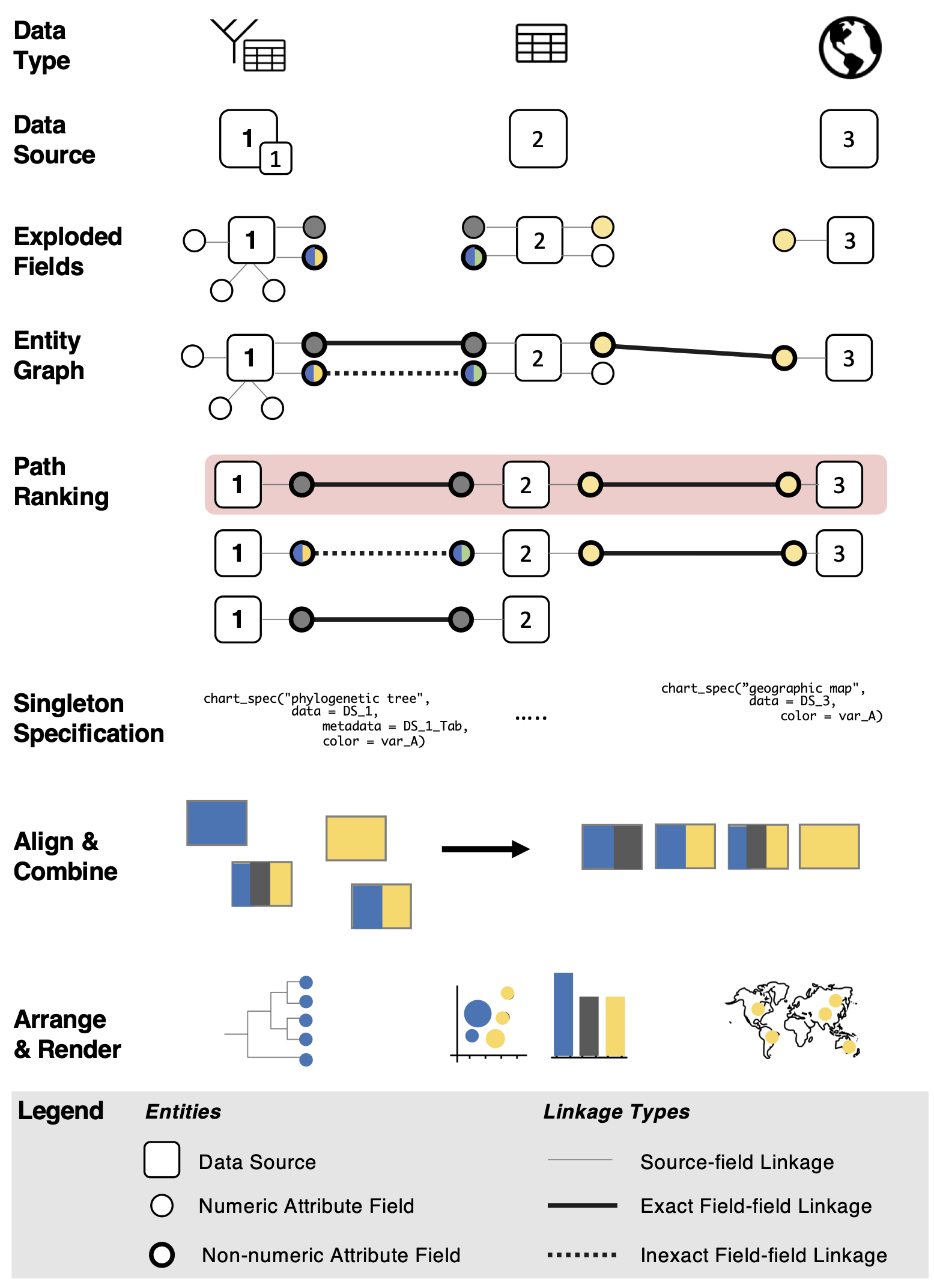}
\caption{
%\textbf{Data Integration and Entity Graph Generation 
\textbf{\SYSTEMNAME\ Overview.} The algorithm is illustrated with sources of three data types: \#1  tree with associated tabular data, \#2 tabular, and \#3 spatial. The exploded attribute fields within these data sources are classified into numeric and non-numeric field types. The similarity of categories between pairs of non-numeric attribute fields is computed with the Jaccard index to establish exact and inexact linkages between data sources. The data sources, their attribute fields, and the linkages between their fields are used to generate an entity graph. The paths of the entity graph that link pairs of data sources are enumerated and ranked according to their link strength, diversity, and total relevance. For each path, in order of rank, partial specifications are generated for individual charts. They are modified to express linkages through aligned spatial axes or color palettes, then arranged into a grid layout and the specifications are rendered into displayable pixels.}
\vspace*{-5mm} 
\label{fig:entity-graph}
\end{figure}

%\rev{In this section we describe our technique for visual data reconnaissance (\TECHNIQUE).} 
Our \SYSTEMNAME\ recommendation algorithm, summarized in~\autoref{fig:entity-graph} and Algorithm~\ref{alg:gevis-overview}, has \tm{six} stages. In the first stage, the algorithm extracts attribute fields from input datasets, a procedure we refer to as ``exploding fields''. In the second stage, the algorithm analyzes all of the exploded fields to detect which ones are shared between datasets, constituting links between them. The algorithm uses these detected shared fields to generate an entity graph encapsulating all connections between the input data. In the third stage, the algorithm analyzes and ranks paths within this entity graph according to their potential relevance to domain experts. The relevance criteria that we propose include broad coverage of input datasets and of different data types, and information from the domain-specific visualization prevalence design space to prioritize visual encodings commonly used in the domain.
% \rev{We emphasize that our technique enables the inclusion of relevance criteria, whatever they may be, rather than having a fixed criteria of relevance. Thus, others are not limited by the relevance criteria we use here.} 
\tm{The fourth stage of the algorithm automatically generates initial partial programmatic specifications for singleton charts. Viable combinations of charts are generated in the fifth stage  by modifying the specifications so that spatial positions or color palettes are aligned to link information according to shared fields. In the sixth stage, charts are arranged into a grid layout that is consistent with the computed alignments, and finally their completed specifications are rendered into boxes of pixels.} 

\subsection{Input Data}

\tm{Our proposed \SYSTEMNAME\ recommender algorithm has four required inputs  ($H$, $D$, $T$, and $A$) and an optional fifth one ($F$).  Two are provided by users. The required input from users is the (frequently) heterogeneous collection of datasets they wish to visualize together ($H$). Our proof of concept implementation supports the following data types for $H$: tabular, tree, genomic, images, spatial (polygons), or network data. Our recommender algorithm will still function if the user provides only a single dataset, or if the datasets they wish to visualize are independent (i.e. have no shared fields). Users have the option to also specify a set of fields ($F_{users}$) that should appear in the final visual encodings, in addition to the automatically detected shared fields that provide linkages between the input datasets.} 

\tm{The other three inputs are not supplied by the user; they are precomputed datasets that we have built into our proof-of-concept implementation. 
% assess the relevance of different visual encodings and generate visualization specifications for display to the user. 
One is a domain-specific visualization prevalence design space ($D$) that is a quantified and aggregated summary of different visual encodings and how frequently they are used, as described in Section~\ref{subsec:formalisms-designspace}.  
%For our work here, $D$ is derived from our prior study that summarizes the visualization practices of genomic epidemiology experts from approximately 18,000 research publications. We provide more details as the specific ways we derived this instance of $D$ in the Implementation Section~\ref{sec:imp}, and here we discuss $D$ in a more general sense. 
Another is a set of predefined template specifications ($T$) for all of the visual encoding types that are built into our proof-of-concept implementation, as discussed in Section~\ref{subsec:spec_gen}, allowing \SYSTEMNAME\ to automatically generate visualizations without any further guidance from users.} The third is manually analysis results $A$, namely the viability matrix for feasible spatial alignments of charts and the list of positionally immutable charts, described in Section~\ref{subsec:align_combine}. 

%\rev{With these inputs provided, our technique will proceed to automatically generated possible coordinated visual encodings. }

%\vspace{-2mm}
\subsection{Exploding Fields from Data Sources}
For each input dataset in $H$, all of its fields are ``exploded''; that is, they are extracted so that they can be easily compared regardless of the original data type of the source. For each input data set, $H_i$,  its fields are denoted $F = (f_1,f_2,...,f_n)$. 
For each field in $F$, \SYSTEMNAME\ also determines whether it is numeric or non-numeric, and the cardinality of the non-numeric fields. 

The automatic extraction of fields required analysis of how attribute fields are stored in each of the supported data types. For tabular data types, these fields are simply the attribute columns. For other data types, these fields can be stored in a variety of different ways, for example, fields in tree data types may be stored within a plain text flat file that also describes the tree's structure. The attribute fields of non-tabular data types could be identifiers that link to an associated dataset, which is often tabular, or may contain important data as a complex concatenated string of data that needs to be parsed. 

\subsection{Generating the Entity Graph}\label{subsec:entity-graph}
Our algorithm conducts a pairwise analysis of all non-numeric exploded fields between data sources to identify potential linkages between data sources. It computes the set similarity for all pairwise comparisons of non-numeric fields between data sources using the Jaccard Index. For a pair of attribute fields A and B, the Jaccard index is $ J(A,B) = {A \bigcap B}/{A \bigcup B} $, where $J(A,B)\in [0,1]$. When $J(A, B)  = 1$ there is an an exact match between all unique categories in the two fields, $J(A, B)  = 0$  indicates no matches, and $0< J(A,B)< 1$ is an  inexact match indicating that some but not all values are in common. 

The Jaccard index is used to derive an entity graph $E$ of data sources and their associated attribute fields, shown in~\autoref{fig:entity-graph}. \SYSTEMNAME\ visualizes the resulting entity graph using a hub and spoke model (\autoref{fig:entity-graph}, \autoref{fig:usage-results}). Rectangular hubs correspond to a data source  (~\hub~) and circular spokes correspond to attribute fields (~\spoke~). The strength of the connection between data sources, via field-field linkages, is shown using thick solid lines for exact matches and dashed lines for inexact matches. Thin solid lines denote linkages between sources and fields.

We limit the linkage check to non-numeric fields %, such as \texttt{Country}, 
because numeric fields lack context to be robustly joined using this set similarity approach: they tend to have so many unique values that matching them to other fields is not straightforward. Extending \SYSTEMNAME\ to handle linkages between numeric fields robustly would be possible and interesting future work, but is beyond the scope of this initial research.

\rev{One immediately obvious limitation of this approach to generating the entity graph is that messy or noisy data may require users to undertake a clean-up process.  Consider two datasets with a field for viral sample IDs that would be expected to have an exact match. If there are typographic errors or missed entries, the Jaccard Index would be sensitive to these issues: rather than an exact match, the linkage would be computed with a lower value ($<1$). However, this situation still provides a useful insight to surface to the expert. The hub and spoke entity graph visualization allows experts to assess the strength of connections and do a targeted investigation into expected connections that failed to materialize. Moreover, since the Jaccard Index is a continuous value, experts can quickly triage their investigations into data quality by, for example, prioritizing those that have lower Jaccard Index values. Even with messy data, our recommendation algorithm will still produce and prioritize visualizations; performance will simply degrade in that fewer coordinated combinations of charts will be generated.}

\subsection{Ranking Graph Paths}\label{sec:alg-pathranking}
After generating an entity graph, the next stage of the algorithm generates and ranks paths between paired groups of data sources. Since data sources may not all be connected to each other, \SYSTEMNAME\ generates paths within individual connected components in the graph. A single connected component of the entity graph could contain anywhere between a single dataset and all input data sources, so the number of paths per component can vary from 1 to beyond ${N_H \choose 2}$, where $N_H$ is the total number of data sources.

\begin{algorithm}
\caption{pathRanking($E$,$D$)}\label{alg:path-ranking}
\begin{algorithmic}[1]
\Require Entity Graph $E$, Design Space $D$
\Ensure List of Ranked Paths, $R_p$
\State rankedPaths $R_p \leftarrow$ new Array
\For {each compontent $C$ in $E$}
    \State $C_p =$ \texttt{enumeratePaths}($E_c$)\Comment{Dijkstra's algorithm}
    \For {each path $P$ in $C_p$}
        \State $P_s \leftarrow$ \texttt{calculatePathStrength}($P$)
        \State $P_d \leftarrow$ \texttt{calculatePathDiversity}($P$)
        \State $P_{vr} \leftarrow$ \texttt{calculatePathVisRelevance}($P$)
        \State add $P_s,P_d,P_{vr}$ as a row to $R_p$
    \EndFor
\EndFor

\State $R_p\leftarrow$ \texttt{columnNormalize}($R_p$)
\State pathScore $\leftarrow$ \texttt{score}($R_p$)
\State sort $R_p$ by pathScore
\State \textbf{return} sorted $R_p$
\end{algorithmic}
\end{algorithm}

\tm{We rank paths in each component of the entity graph with a \textbf{relevance rank} that incorporates three metrics that we developed:   the strength of the connections between data sources (\textbf{link strength}), the diversity of data types (\textbf{data type diversity}), and the cumulative relevance of visual encodings that could be generated from those data types (\textbf{visual encoding relevance}). The visual encoding relevance metric is how we incorporate domain-specific visualization practices into the path ranking criteria, specifically the genEpi visualization prevalence design space available from the GEViT study~\cite{gevit2018}.}

%% TM: this argument belongs later, in Results! 
%The calculation of relevance is nuanced and based upon the characteristics our understanding of genEpi domain expert practices. But more generally, the \TECHNIQUE\ technique we present here is capable of flexibly adapting to different metrics that we ourselves or others wish to explore in future. For the purposes of our genEpi application, however,  paths that have strong links between data sources of different data types and that can produce several types of visual encodings that are relevant to genomic epidemiology experts are highly ranked by \TECHNIQUE}. We developed these ranking metrics so that \TECHNIQUE\ produces a broad overview of the data using data visualizations that are relevant to genomic epidemiologists. 

We summarize our approach to path ranking in Algorithm~\ref{alg:path-ranking}. %Below we provide detailed calculations of our ranking metrics. 
The \textbf{link strength ($P_s$)} of a path is the normalized sum of edge weights ($e_i$):
\begin{equation}
    P_s = \frac{\sum_{i=1}^{N_{e}} e_i}{N_{e}}
\end{equation}
where $N_{e}$ is the total number of edges on that path and $e_i$ is the Jaccard index of the $i^{th}$ edge, and finally $P_s\in[0,1]$. A value of $P_s = 1$ indicates that data sources are linked entirely by exact matches, whereas $P_s = 0$ indicates that there is only one data set in that path. 

\textbf{Data type diversity ($P_d$)} is calculated by summing the number of unique data types in a path. Each data source is classified as belonging to one of the supported data types; in the implementation, this calculation is done at runtime when the user loads the input data into \SYSTEMNAME . A diversity value of 1 indicates that every data source appearing on the path is the same data type, for example tabular data. The maximum possible diversity value of $N_H$, the total number of input sources, would indicate that each data source is a different data type. The objective of the diversity metric is to include a number of different data sources, but also to prioritize visualizing data across a variety of data types.

Finally, we compute the \textbf{visual encoding relevance ($P_{vr}$)} of a path by summing the relevance scores of the unique visual encodings that can be produced from the different data sources along the path. We have pre-computed a scaled visual encoding relevance ($R'$) using the results of our prior analysis of visualization practices in genomic epidemiology~\cite{gevit2018}. The $R'$ value is derived from a quantification for visual encoding ($V$) usage within a domain-specific visualization prevalence design space ($D$), where $R' \in [1,10]$.   We compute R as follows:

\begin{equation}
   R = \sum_{y}^{Y} \sum_{n=1}^{N_y} V_{i,n}\times w_y
\end{equation}

where $V_i$ is an individual visual encoding in the form of a chart (i.e. scatter chart, map, phylogenetic tree). A simple relevance metric score would just quantify how often some visual encoding ($V_i$) is used in some design space ($D$); for the prior GEViT study~\cite{gevit2018}, this number captures how many papers a chart type appears in. 

We sought to give greater weight to more recent visual encodings, reasoning that best practices were likely to evolve over time. Thus, we create a penalty term ($w_y$) that down-weights counts from older years, a possible computation since the GEViT VPDS includes information about the year of use $Y$. % study, it gives less weight to chart types published in older papers. 
This approach sums the occurrence of individual visual encodings, while giving greater weight to visual encodings that domain experts used most recently.

For consistency across evolving design spaces, we re-scale the visual encoding relevance ($R$) as follows:

\begin{equation}\label{eq:scaled_rel}
R' = \left( \frac{V_i}{max(R)}\right) \times 10
\end{equation}

The above equation identifies the visual encoding ($V_i$) that is used most frequently in this design space ($R$)). All the visual encodings in the design space are scaled as a fraction of $R$. We include a scaling factor of 10  for interpretability and ease of computing; other scaling factors could have been used. The scaled relevance score ranges between 1 (least common) and 10 (most common). It is designed to produce non-linear quantitative values to emphasize the relative importance of different types of visual encodings, not simply an ordering of the elements. 

For example, a phylogenetic tree is the most used visual encoding in genEpi and the next most common visual encoding is a bar chart. Our scaling produces scores of 10 and 4 respectively for these charts, rather than scores like 10 and 9, to capture the relative importance of a phylogenetic tree compared to all other encodings.

We  calculate the total relevance of a path by calculating $P_{vr}$ as follows: 
\begin{equation}
    P_{vr} = \sum_{1}^{I_p}max(R')
\end{equation}

where $I_p$ is the total number of datasets in the path and $1\geq I_p \leq N_H$. In some instances, a data type may map to multiple different possible encodings,% (see~\autoref{fig:mapping-chartypes}), 
for example with tabular data that can map to various statistical chart types (bar chart, scatter chart, etc.). In this case our algorithm only considers the highest value encoding ($max(R')$) that a dataset maps to. Other data types only link to one visual encoding, for example, spatial polygons can only be drawn as a map; in this case, there is only a single value of $R'$. Consequently, $P_{vr} \in [1, max(R_D')\times I_p]$, where $R_D$ is the visual encoding with the highest value in the entire design space. As a concrete example, a phylogenetic tree is the most important visual encoding in our existing design space and would have $R' = 10$. If the user supplies five input datasets, each of which are phylogenetic trees and are all connected to each other via a common sample ID, then the value of $P_{vr} = R'\times 5 = 10 \times 5 = 50$. The $P_{vr}$ metric enables our technique to prioritize data types that are likely to produce visual encodings the user may care most about, given the domain-specific information captured within a VPDS.  

%\SYSTEMNAME\ computes the relevance of each individual visual encoding that can be produced from the input data sources, again relying on the mapping between data source and data types to do so (~\autoref{fig:mapping-chartypes}). Finally, \SYSTEMNAME\ simply sums relevance of individual visual encodings. 

As a final step, we rank normalize the values of $P_s$,$P_d$,$P_{vr}$, such that $P_s$, $P_d$, $P_{vr} \in [1,P]$, where 1 indicates that highest ranked and $P$, the number of paths, is the lowest ranked value. The final importance score for a path is established by summing the normalized values of path strength, diversity, and encoding relevance: $pathScore = { P_s + P_d + P_{vr}}$, where $pathScore \in [3,3P]$.

\subsection{Generating Singleton Specifications}\label{subsec:spec_gen}
\begin{algorithm}
\caption{generateVisSpecs($R_p, M,F',F_{user},A, T$)}\label{alg:genVis}
\begin{algorithmic}[1]
\Require rankedPaths $R_p$, field metadata $M$, Fields $F$, User Specified Fields $F_{users}$, Encoding Templates $T$
\Ensure array of chart combinations $V_C$ 
\State $V_C \leftarrow$ new array
\For{\textbf{each} path P$\in R_p$}
    \State ($F_p,M_p$)  $\leftarrow (F,M)\in (P \subset F = field)$ \Comment{Filtering}
    \State sortDescending($F_p$, by degree) \Comment{Sorting}
    \State $V \leftarrow$ \texttt{generateSingleChartSpecs}($F_p,F_{user},T$)
    \State $V_c \leftarrow$ \texttt{generateCombinationSpecs}($V, A$)
    \State  $V_C \leftarrow [V_C,V_c]$
\EndFor
\State \textbf{return $V_C$}
\State
\Function{generateSingleChartSpecs}{$F_p,F_{user},T$}
\State $F'\leftarrow[F_{user},F_p]$ \Comment{Sorted order by importance}
\State $V \leftarrow$ new array
\For{\textbf{each} chart $C$ in $T$}
    \For{\textbf{each} encoding slot $C_e$ in $C$}
        \State \textbf{try:} $C_e \leftarrow f$ where $f \in F'$
    \EndFor
    \If{$(\forall C_e\in C) \neq$ \texttt{NULL}}
        \State $V \leftarrow [V,C] $
    \EndIf{}
\EndFor
\State sort $V$ by chart relevance
\State \textbf{return} $V$ 
\EndFunction
\State
\Function{generateCombinationSpec}{$V$, $N=5$}
\State $V' \leftarrow [V_1...V_N]$
\State $V_{spatial}' \leftarrow$ \texttt{whichSpatiallyAlign}($V'$)
\State $V_{color}' \leftarrow$ \texttt{whichColorAlign}($V'$)
\State \textbf{return} [$V_{spatial}'$,$V_{color}'$]
\EndFunction
\end{algorithmic}
\end{algorithm}

\begin{figure}[tb]
\centering % avoid the use of \begin{center}...\end{center} and use \centering instead (more compact)
\includegraphics[width=\columnwidth]{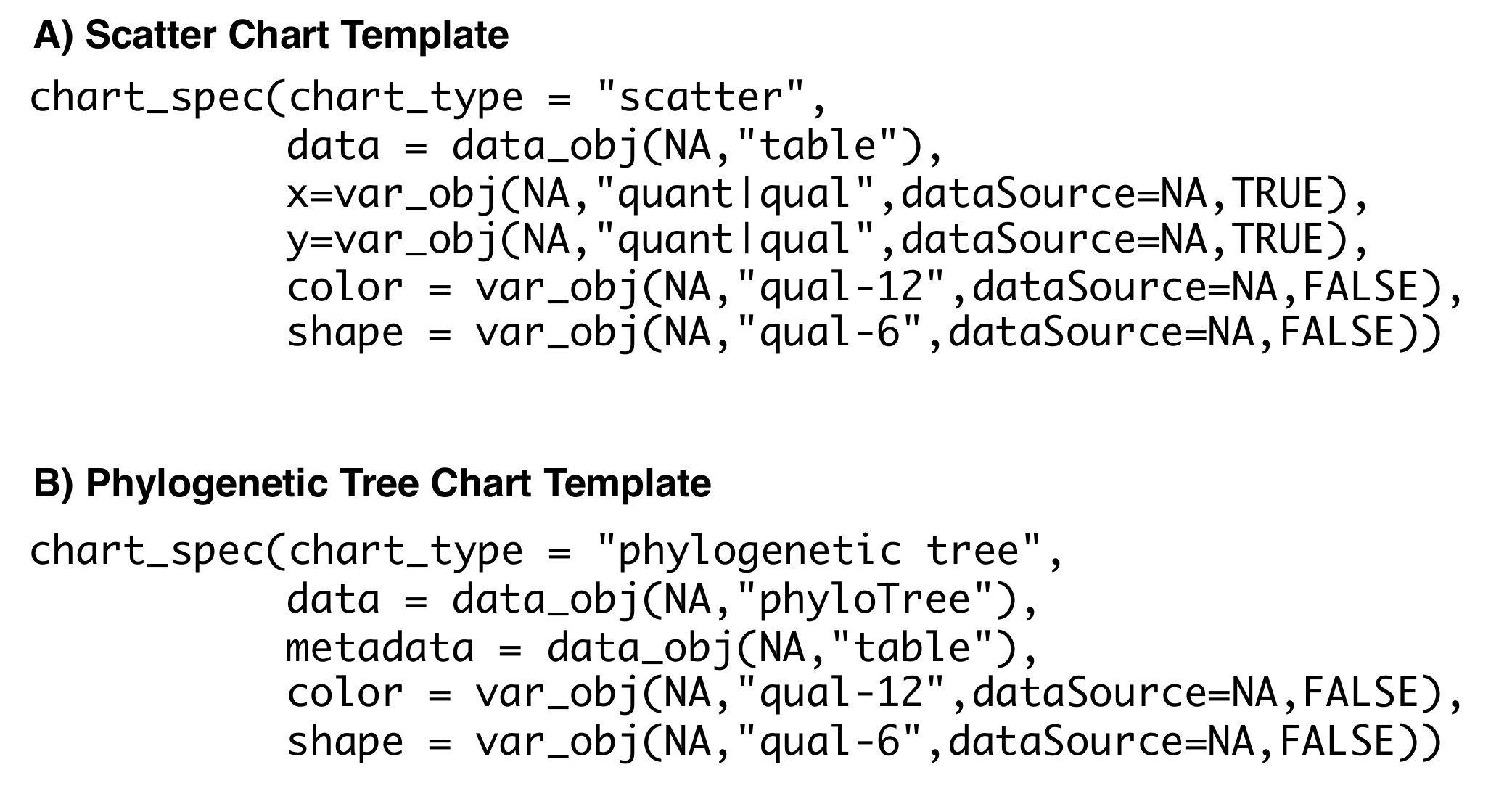}
\caption{Chart Templates. These initial partial specifications for chart templates are pre-defined and internal to \SYSTEMNAME\, not exposed to the user. Examples: (a) Scatter chart. (b) Phylogenetic tree.}
\label{fig:vis-template}
\end{figure}

Beginning with the highest ranked path ($pathScore = 3$) \SYSTEMNAME\ generates programmatic specifications for coordinated and combined visual encodings in two passes, as described in Algorithm~\ref{alg:genVis}. The first pass generates a specification for each singleton charts using a pre-defined set of chart templates, one for each supported chart type. At run time, chart templates are converted to declarative specifications that facilitate layout and rendering of charts. 
It would have been possible to serve the same purpose with a bespoke query language such as CompassQL, as used by Draco~\cite{Moritz2019} and Voyager~\cite{Wongsuphasawat2016,Wongsuphasawat2017}, or the similar query language used by ShowMe~\cite{mackinlay_show_2007}. It would also be possible to use a grammar, as ggplot does~\cite{ggplot}. However, both of these approaches require substantial input from the user. We instead chose the template approach to eliminate that overhead, as well as for ease of debugging and to allow flexible experimentation.

A chart template $T$ contains slots of the chart type, data source (and if applicable metadata), and visual encoding channels (x, y, color, shape). Data and encoding slots contain additional parameters that are used to verify the suitability of some dataset or field to generate the encoding. Data slots contain parameters for the data source name (NA if unassigned) and data type (table, tree, network, spatial, image). Encoding slots contain parameters for field name (NA if unassigned), field type constraints (numeric or non-numeric), field data source, and whether that field is required to generate the encoding or optional. Data source name, field name and field data source are dynamically assigned at run time.  
Two examples of templates are shown in \autoref{fig:vis-template}. 

The constraints on a visual encoding slot vary by the channel type, following established perceptual guidelines~\cite{cleaveland1984, Mackinlay1986, mackinlay_show_2007, Moritz2019}. The positional (x,y) encodings slot can be filled by either numeric fields or non-numeric fields with high cardinality. Color encoding slots are constrained to non-numeric fields that have fewer than 12 categories. Size encoding slots are limited to numeric fields.

At run time, \SYSTEMNAME\ sorts fields (nodes) in a path according to their degree of connectivity. Field nodes with a degree of 2 or more connect multiple datasets and are given higher priority than non-connecting fields when assigned to slots in chart templates. The user has the option to also specify fields that should appear in the final visual display and these are given the highest priority of all.  \SYSTEMNAME\ attempts to assign fields in order of importance to appropriate visual encoding slots of chart templates. If a field does not meet the constraints for a encoding slot, our algorithm tries to fit it to another encoding slot or leaves the field unassigned. Only charts that have all of their required encoding slots assigned to a field move forward to the next pass where combination specifications are generated.

\subsection{Aligning and Combining Charts}\label{subsec:align_combine}

\tm{In the next stage, \SYSTEMNAME\ combines singleton charts together. Each path will result in a combination of multiple charts that have been explicitly coordinated to show linkages across multiple data sources, where the shared fields appear in multiple charts to express the linkage visually.  At a minimum, a single chart is produced per each component of the graph.}

\tm{Our recommender algorithm is intended to leverage existing software layers designed to render charts, such as D3~\cite{bostock2010}, Vega~\cite{2017-vega-lite}, or ggplot~\cite{ggplot}. However, the problem of leveraging existing software when automatically coordinating static charts is not trivial. Once a chart is rendered into a small box of pixels, no aspects of its visual encoding such the colors or spatial position of marks can be updated or changed. We solve this coordination problem with a declarative approach that we call \textbf{gradual binding}: only partial specifications are initially generated, which are then modified in discrete stages. After a specification is finalized and complete, it is passed along to a charting library for rendering. This gradual approach allows \SYSTEMNAME\ to use extremely minimal initial specifications for the singleton charts and for the type of combination. The algorithm then automatically derives additional specifications that enforce the necessary consistency within each chart to instantiate the requested combination, so that all of the rendered boxes of pixels can simply be concatenated to achieve the desired result.} 

%Thus, in the previous stage, \SYSTEMNAME\ produces only initial partial specifications for individual charts. 
%One additional step is required to coordinate the charts for display. 

\tm{The three chart combinations supported in \SYSTEMNAME\ are spatial alignments, where two or more charts share a common field in the positional encoding slot; color alignments, where the same field is assigned to a color encoding slot for two or more charts; and unaligned combinations, where there is no visual linkage. These combinations are not mutually exclusive: all three may simultaneously occur. 
%\textbf{Spatially aligned} chart combinations share a common field in the positional encoding slot. 
The GEViT study~\cite{gevit2018} also identified small multiples as a common combination within genEpi.  Our algorithm does not support that combination type, to keep the scope of \SYSTEMNAME\ tractable; handling coordination seamlessly between all three of these combination types is an open research challenge, that we leave for future work.}

%The primary mechanism for facilitating coordinated combinations is shared fields that can be specified for use in visual encodings to create a visual linkage. 
%Singleton charts can be  modified to accommodate spatial alignments through reorienting operations such as rotations. We have also identified additional constraints that limit the types of charts that may be combined.

\textbf{Color aligned} combinations will automatically apply a common color palette for shared attribute fields across multiple chart types. Color alignment is only facilitated between charts that have a common field assigned to their color encoding. At run time, \SYSTEMNAME\ recognizes this commonality and enforces a common color palette across the individual charts. Although the automatic coordination of color between views is supported in many existing systems for small multiples, most previous systems do not support this capability across multiple and diverse chart types. 
%Our work extends beyond existing systems that primarily support automatic coordinated color alignments only for small multiples, not multiple chart types. 

\tm{For \textbf{spatially aligned} combinations, \SYSTEMNAME\ needs to ensure that charts have a common spatial axis along either the horizontal or vertical direction. There are two considerations: whether it is feasible to spatially align charts the charts at all, and how to orient them.}

\begin{figure*}[th!]
  \centering 
  \includegraphics[width=\textwidth]{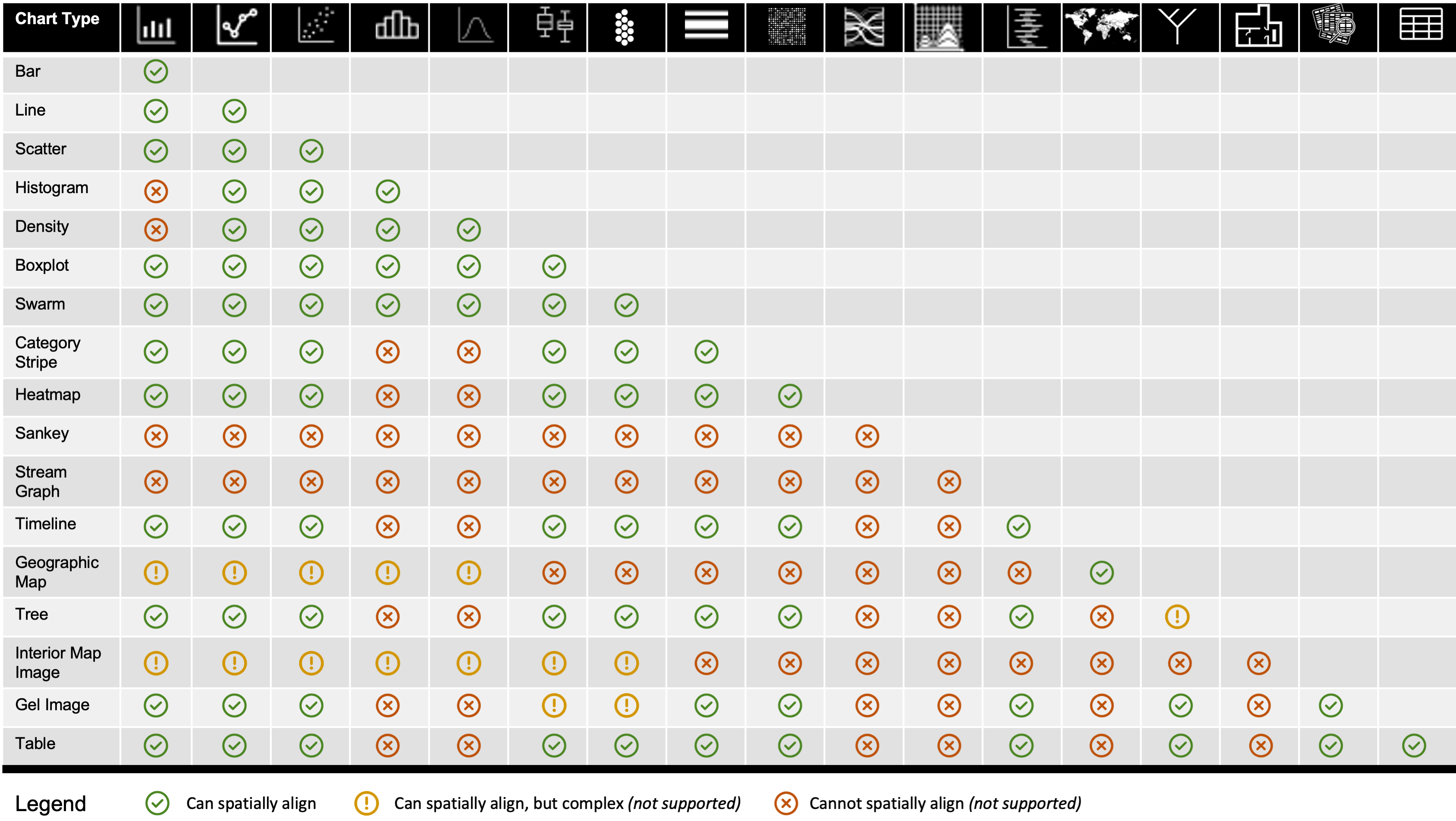}
  \caption{ \textbf{Compatible chart types for spatial alignments.}}
  \label{fig:sp_compatible}
\end{figure*}

\tm{Determining a shared axis of alignment is straightforward when combining charts that all use Cartesian coordinates, with a shared field either the x or y axis, as is the case with many common statistical charts. However, domains with a diversity of chart types, including genEpi, require more extensive consideration in finding a shared axis. The solution we propose in \SYSTEMNAME\ is to manually analyze the full set of charts used in the domain design space (through discussions between the authors), to determine viable combinations where it may be possible to establish a shared axis. The result of our analysis is a viability matrix, shown in~\autoref{fig:sp_compatible}, that is used by the \SYSTEMNAME\ algorithm to automatically determine whether spatial alignments are viable. 
We assigned each combination into one of three categories: possible and straightforward to programmatically support, possible but not easy to programmatically support, and not possible. We made these determinations of compatibility based upon the types of axes (numeric and non-numeric) and coordinate systems of the chart. For example, a histogram contains two axes (x and y) that both have numeric data, and thus is theoretically compatible with other charts types that have numeric axes. Our current \SYSTEMNAME implementation only supports possible and programmatically feasible spatial combinations.}
%The recommender algorithm has already determined whether each field in the input data is numeric or non-numeric in earlier stages of the algorithm, so this information can be incorporated into the alignment computation in this stage.

%, further automation of this processes a fruitful area of future work, but for a proof-of-concept implementation it still provides a large and valid range of multi-chart chart combinations.

% (our \TECHNIQUE\ technique makes the final determination by examining the data domains for candidate alignment fields). 

The square matrix in~\autoref{fig:sp_compatible} addresses the 20 chart types handled in \SYSTEMNAME\, out of the 25 contained in the GEViT design space. Only 17 rows and columns are present because we collapsed several structurally identical chart types into the same rows: Tree represents both phylogenetic and dendrogram trees; Geographical Map includes choropleth maps; and Tables includes genomic alignments. We chose to omit the 4 chart types of images, networks, pie charts, and Venn diagrams upon determining that these charts are unlikely be part of spatially aligned combinations because they lack a horizontal or vertical axis to facilitate such combinations. For example, a pie chart uses a radial coordinate system, whereas nearly all genomic charts use Cartesian co-ordinates. For network diagrams, the specific coordinates of the points often have no spatial meaning and can occlude one another if projected either horizontally or vertically; Venn diagrams have similar problems. The pixel space within images is sufficiently unrelatable to other genEpi data that it is typically not informative to spatially align other charts to them (with the exception of pulse field gel electrophoresis images and a few other specialized cases). 

%derives an ordering and orientation for each of the individual charts. 

\tm{While some charts are straightforward to reorient, other chart types should not have their positional coordinates altered because it would inappropriately distort the information they contain. We again propose a solution based on manual analysis of the charts used in a domain design space, and then using those results programmatically within our recommendation algorithm. We analyzed all of the chart types that occur in the GEViT design space and identified the following 3 types as positionally immutable: trees, geographic maps, and images (with some exceptions). We refer to the positionally immutable charts as \textit{lead charts} and the others as \textit{support charts}. The \SYSTEMNAME\ algorithm is constrained to generate specifications with only one or zero lead charts; if no charts in the specification are positionally immutable, a lead chart is randomly selected. All of the support chart specifications are modified to rotate each one to share a common axis with the lead chart. A common scale is then automatically generated and applied to all charts based upon the scale of the lead chart.} 

% determined at run time when establishing the set of spatially alignable charts. 

%Spatial and color alignment are not mutually exclusive. Charts can be simultaneously spatially and color aligned, as shown in~\autoref{fig:synthetic_data_archiecutre}. 
\tm{Finally, \SYSTEMNAME\ may also generate unaligned combinations, where there are no shared fields between visual encodings. This situation can arise when two datasets are linked by a common domain (for example, a data ID), but this variable is not included in the visual encoding. The purpose of aligned combinations is to get different 'snapshots' of the same data, similar to Voyager or ShowMe, but to still ground those snapshots in data source relationships. Since our prior study found this combination to be less common, these paths are not likely to be the highest ranked but may appear as lower ranked ones.} 

\tm{The gradual binding process concludes when complete chart specifications have been derived that fully adhere to the indicated combination types. Any stylistic defaults built into the underlying charting libraries will be inherited, unless they have been explicitly overridden.} 

%\label{subsec:feasiable_align}

%Individual singleton charts are generated first and viable spatial and color aligned combinations are determined at run time (detailed in Section~\ref{subsec:layout_render}).

\subsection{Arranging and Rendering Boxes of Pixels}\label{subsec:layout_render}
%\subsection{Layout and Display}\label{subsec:display}

In the final stage of the algorithm, arranging and rendering the charts is straightforward because the finalized specifications emerging from the previous stage have harmonized alignments and orientations, as the outcome of the gradual binding. 
The algorithm assumes access to a charting library that will take a complete specification and return a rendered box of pixels. The only further requirement is to arrange the boxes of pixels that represent rendered individual charts into a grid. 
By default, \SYSTEMNAME\ creates a 2$x$3 grid; that configuration is dynamically modifiable at run time. The arrangement depends upon the types of combinations that are feasible and any user specifications to modify the number of charts that compose a view. % Spatially aligned combinations, if they are feasible, exclusively occupy the top row of the grid. Color aligned combinations and other charts that are neither spatially or color aligned tend to be placed second row. 

The final output of the recommender algorithm is a single large box of pixels, showing the complete combination of static visualizations with linked information content.
%, as shown in~\autoref{fig:usage-results}. 

\section{Implementation}\label{sec:imp}

\tm{We now describe the proof-of-concept implementation of our recommendation algorithm. It provides evidence for the viability of our approach, and allows us to conduct a preliminary evaluation of its utility.} 

%\rev{In this section we present a proof-of-concept implementation our \TECHNIQUE\ technique as \SYSTEMNAME. We describe out implementation as a proof-of-concept, rather than a system, because there exist additional technical and theoretical challenges that need to be resolved before we can develop a robust system. We describe those challenges in below in sections~\ref{sec:background} and~\ref{subsec:feasiable_align}.  Despite these limitations, a tangible implementation of a technique is a useful mechanism to assess that it's ideas are viable. Moreover, for techniques that produce visualizations human feedback on its outputs are important to formative assessments, and this also requires some tangible interface. The term proof-of-concept succinctly encapsulates these concerns and still enables us to conduct preliminary evaluations.}

\subsection{Language and Environment}
\tm{We chose to implement \SYSTEMNAME\ as a package in the R programming language.  We chose that language to fit into the workflows of genEpi experts: they routinely use R to perform statistical analyses, and integration within it allows them to use that environment to further filter or refine their analysis. The static combinations of charts that are created by \SYSTEMNAME\ are also usable for communication with others in addition to supporting the data recon process, since PDF reports remain the primary medium of communication between genEpi experts~\cite{peerJ2018}.} 

\tm{We thus build on ggplot~\cite{ggplot} as the charting library layer for many of the statistical charts, and a variety of other R packages such as ggtree~\cite{ggtree2017} for more specialized charts. Figure~\ref{fig:sp_compatible} enumerates the set of charts that are supported in our current implementation.} 

Our package code and evaluation data is online: \url{https://github.com/amcrisan/GEViTRec}

%The \SYSTEMNAME\ implementation contains methods to load heterogeneous data, generate the entity graph, generate visualizations specifications, and finally lay out and render the coordinated combinations of views to the display. 

%The first dataset we derived is a domain prevalence design space for the calculation of the relevance metric (Section~\ref{alg:path-ranking}). The second dataset we produce was a manually curated list of viable spatial chart combination in genomic epidemiology (Section~\ref{subsec:spec_gen}), \autoref{fig:sp_compatible}. In the prior section, we describe how these data are used by our \TECHNIQUE\ technique, and here we describe their derivation for use by \SYSTEMNAME.

\subsection{GEViTRec API}
%\ac{Added this back from CHI, was absent from KDD submission}

\tm{The Application Programmer Interface (API) for the \SYSTEMNAME\ package is designed for extreme simplicity, in keeping with the goal of a  recommender system suitable for data recon. It is intended to be run from an R environment as a set of functions, which was sufficient to demonstrate the concept; it could be transformed into a standalone interactive application as future work.} 
% interactive Shiny application interactivity could be added later if \SYSTEMNAME\ were to be further developed into a standalone system. 

The API has functions to help users load heterogeneous data, perform data integration, generate specifications for chart combinations, and render those specifications as displayable plots. %\autoref{fig:usage-results} shows example results. 

\texttt{input\_data} is a common interface for loading different data types into R. This function requires as input the location of the data source on disk and the data type. If applicable, associated data for non-tabular data types can also be loaded. We have developed a series of data-type-specific functions that will load  and store a dataset in a standardized format.

\texttt{data\_linkage} takes a collection of datasets and explodes attribute fields from different data sources, finds linkages, and creates the entity graph. The \texttt{view\_entity\_graph} command will display that graph. Users can also view a metadata table for the different data sources and their attribute fields. 

\texttt{get\_spec\_list} performs the computations on the entity graph. It ranks paths and processes them in order of rank (highest to lowest) to generate specifications for charts.

\texttt{plot\_view} takes a list of chart specifications, renders them, and arranges the resulting pixels together into a view: a single large box of displayable pixels, containing multiple coordinated static charts. A user parameter indicates which of the views to display. 

Separating the data integration and specification computation into separate functions allows the optional display of the entity graph data structure itself, if desired by the user. 

In instances where an entity graph contains multiple disconnected components, our current implementation limits the total number of views per component to ten; that is, we assemble views from up to ten paths within a single component. Our implementation also limits the number of chart types per coordinated static combination to five, which are selected based upon their relevance scores. However, these limitations are modifiable. 

%\subsection{Architecture}

% \begin{figure}[h!]
%  \centering
%  \includegraphics[width=0.95\columnwidth]{../figures/implementation_overview.png}
%  \caption{GEViTRec Implementation Overview}
%  \label{fig:implementation}
% \end{figure}

% Figure~\ref{fig:implementation} illustrates the software architecture of our implementation. The initial specification and the display 

\section{Results}
\label{sec:results}

\begin{figure*}[th!]
\centering % avoid the use of \begin{center}...\end{center} and use \centering instead (more compact)
\includegraphics[width=0.90\linewidth]{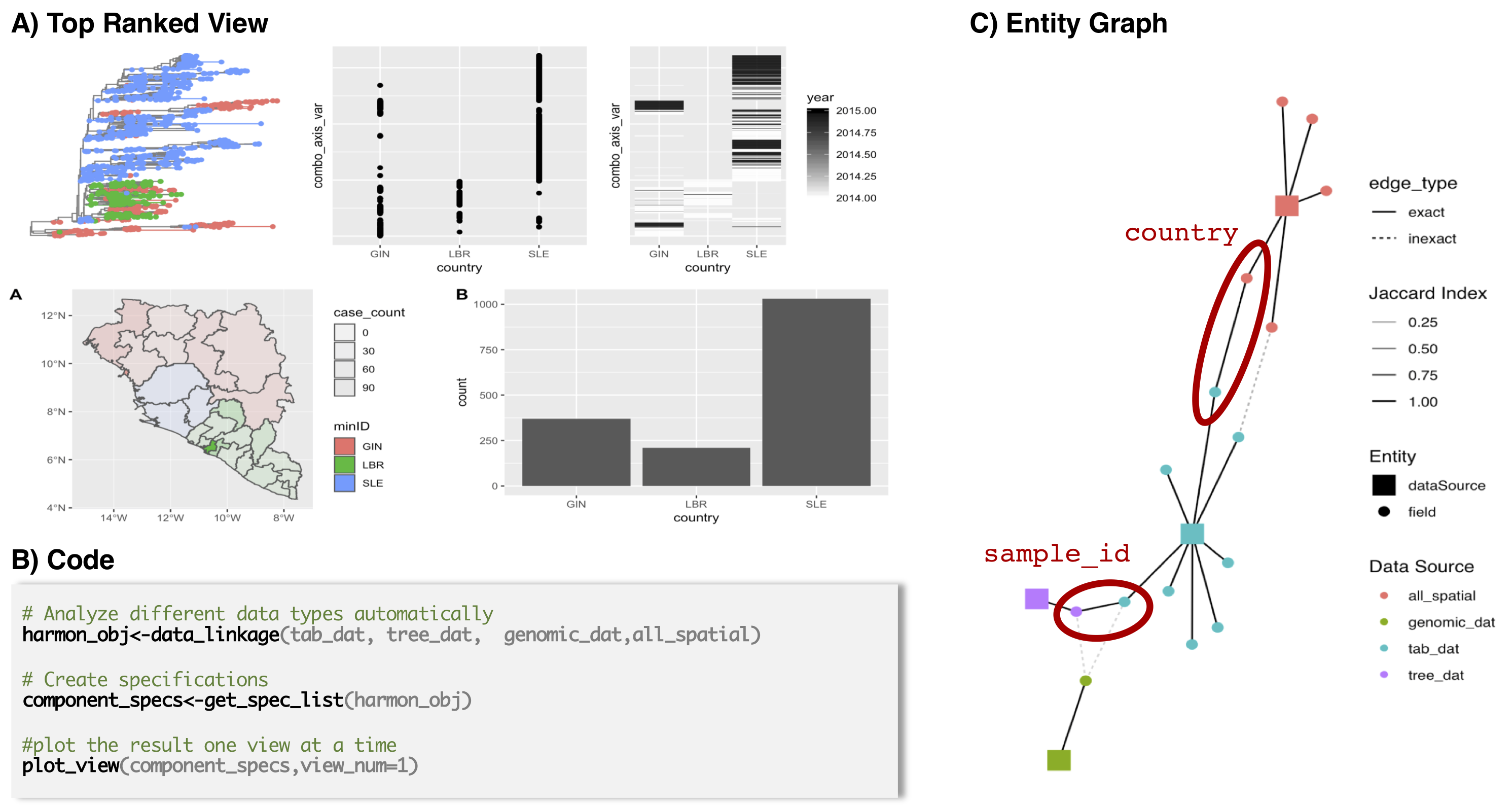}
%\caption{Ebola Dataset. A) Top ranked view. B) \SYSTEMNAME~code. C) Entity graph. \SYSTEMNAME\ generates all views in 35 seconds.}
\caption{\tm{\textbf{\SYSTEMNAME~results: real Ebola outbreak data.} A) The highest ranked view generated by \SYSTEMNAME contains five charts of different types, featuring a spatially aligned combination across the three top-row charts and color alignment between the tree on the to prow and the geographic map on the bottom row.  B) The \SYSTEMNAME\ code required to generate this view. C) The entity graph generated by \SYSTEMNAME, which can be displayed on demand.}}
\label{fig:usage-results}
\end{figure*}

%%%%%%%%%%%% MINCOMBINR %%%%%%%%%%%%%%%%%%%%%%%%%%%%%
% \begin{figure}[h!]
%  \centering % avoid the use of \begin{center}...\end{center} and use \centering instead (more compact)
%  \minipage{.8\columnwidth}
%     \includegraphics[width=\linewidth]{../figures/synth-results-supp.png}
% \endminipage\hfill  
%     \minipage{.2\columnwidth}
%  \includegraphics[width=\linewidth]{../figures/chart_layout_diagram.png}
% \endminipage
%  \caption{\SYSTEMNAME~results with synthetic data. B) Synthetic Dataset Results. C) Coordinated Combinations Summary.}
%  \label{fig:synthetic_data_archiecutre}
% \end{figure}

\begin{figure}[th!]
\centering
\includegraphics[width=0.90\columnwidth]{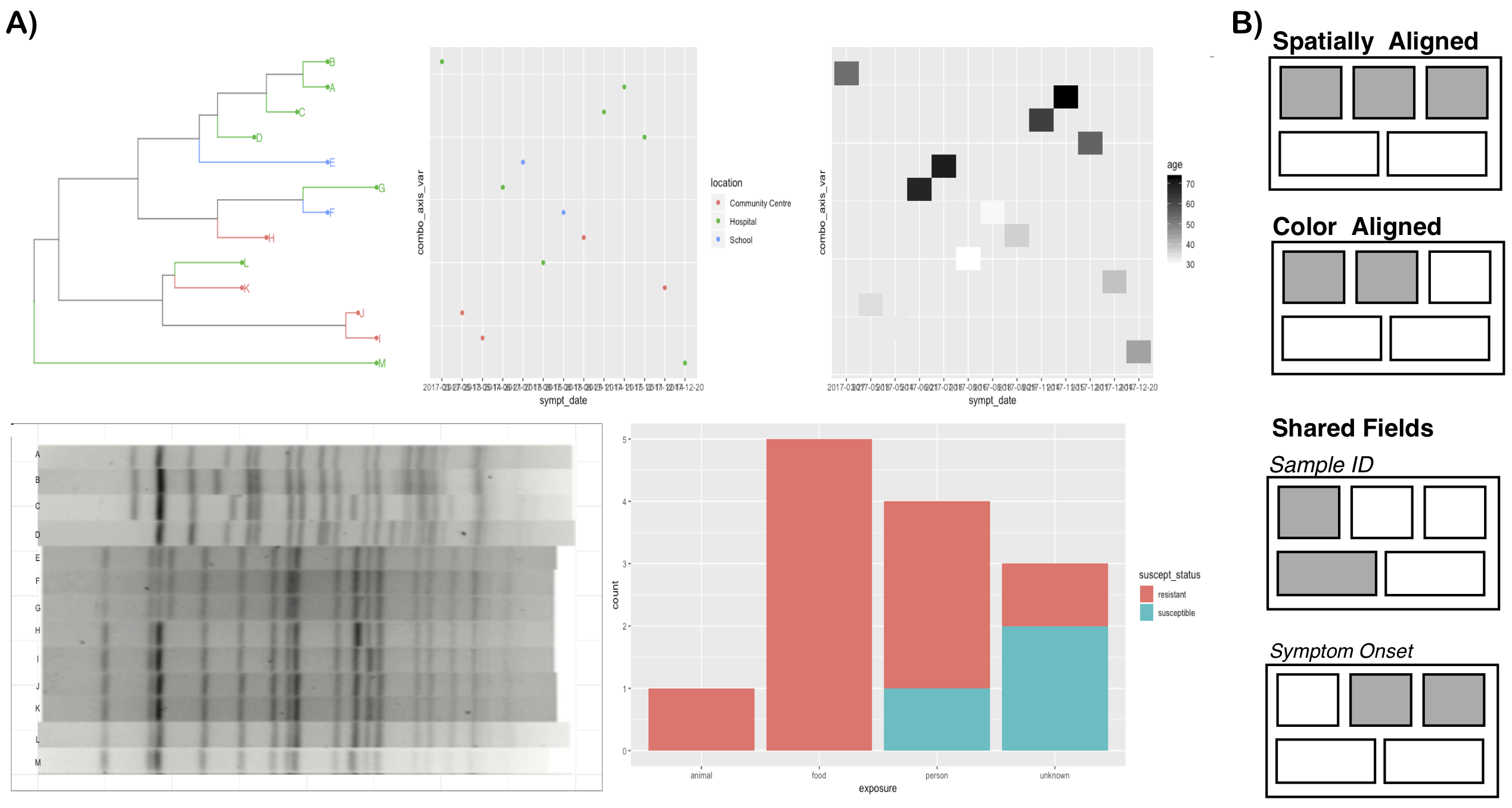}
\caption{\tm{\textbf{\SYSTEMNAME~results: synthetic genEpi data.} A) The fourth ranked view generated by \SYSTEMNAME\ for simple dataset with 13 samples. B) Diagram depicting the types of alignments between the five charts.}}
\label{fig:synth-data}
\end{figure}

We use both real and synthetic genEpi data to assess \SYSTEMNAME's performance. In this section, we briefly describe these datasets and show one example chart combination for each. We provide detailed vignettes of the recommender outputs in the supplemental materials, with a walkthrough of each of the generated combination views. 

\subsection{Real Ebola Outbreak Dataset}

\tm{The real genEpi dataset collection is publicly available data from the 2014-2016 Ebola outbreak~\cite{Dudas2017}. It includes a phylogenetic tree for roughly 1610 Ebola virus genomic samples, one from each affected person. The spatial dataset has case counts for the affected nations. A tabular dataset contains additional information for each sample. 
The results shown in Figure~\ref{fig:usage-results} were generated in approximately 30 seconds using a 2017 MacBook Pro.} 

\tm{Figure~\ref{fig:usage-results}A shows the highest ranked view created by \SYSTEMNAME, a 2$x$3 grid of 5 charts.  
The top row shows a spatially aligned combination of the phylogenetic tree, scatter chart, and heatmap. The phylogenetic tree was chosen by the algorithm to be the lead chart, so the vertical ordering of the phylogenetic tree leaves (representing affected people) is shared with the scatter chart encoding the country of the case, and with the heatmap that encodes the temporal progression of the outbreak. This spatial alignment is explicitly indicated by the \texttt{combo\_axis\_var} label stating that the charts share a common y-axis and can be read together horizontally across the row. The bottom row has two charts showing case counts, one with a choropleth map that contains fine-grained information about cases at the region level within each country, and the other with a bar chart aggregating cases into the country level. The geographic map in the bottom row is color aligned with the phylogenetic tree in the top row, where the shared field is \texttt{country}.}  

\tm{Figure~\ref{fig:usage-results}B shows the extremely concise \SYSTEMNAME\ code required to generate this view. 
The entity graph that was automatically derived from the separate input datasets is shown in Figure~\ref{fig:usage-results}C, 
as displayed by the \texttt{view\_entity\_graph} function.}

\subsection{Synthetic Simple Dataset}

\tm{The small synthetic dataset that we created as a simple example to demonstrate the capabilities of our system features four fabricated datasets with 13 samples in each: tabular data, a phylogenetic tree, genomic data, and a pulse field gel electrophoresis (PFGE) image. The results shown in Figure~\ref{fig:synth-data} were generated in approximately 17 seconds, again by a 2017 MacBook Pro.}  

\tm{The fourth-ranked view generated by \SYSTEMNAME is shown in Figure~\ref{fig:synth-data}A, alongside a diagram that we manually created to illustrate the types of coordinated combinations in that screenshot in Figure~\ref{fig:synth-data}B. The three spatially aligned charts in the top row all share the same vertical ordering for samples, dictated by the phylogenetic tree. The left tree and the middle scatter chart are color aligned through the \texttt{location} field, and the middle scatter chart and the right heatmap showing peoples' ages have a shared field, the date of symptom onset, which has been chosen as the horizontal axis for both. The sample ID shared field appears in both the tree on the top left and the PFGE image on the bottom left, but these two charts are not spatially aligned: the vertical ordering is different. Neither trees nor gel images can be reordered, so \SYSTEMNAME\ did not attempt to align them spatially. Notably, the stacked bar chart on the left showing exposure vectors and susceptibility status does not share fields with any of the other charts. They occur within the tabular dataset, which does have shared fields with the other datasets, so it is within the same connected component of the entity graph. It is an example of how an unaligned combination can usefully provide additional information that may be of interest during data recon, which should not be limited to only charts that can be aligned with each other.} 

\section{Comparison to Existing Tools}
\label{sec:eval-systems}
We compare the visual encodings generated by  \SYSTEMNAME\ to those from the Voyager~\cite{Wongsuphasawat2016}, Draco,~\cite{Moritz2019} and ShowMe~\cite{mackinlay_show_2007} visualization recommender systems, as well as to
two state-of-the-art genEpi tools that were manually curated by humans,
Nextstrain~\cite{hadfield_nextstrain:_2018} and Microreact~\cite{Argimon2016}. 
%There are no directly comparable systems to \SYSTEMNAME\ (see Section~\ref{sec:related-recommender}). We establish a baseline for performance, conduct formative evaluations using a synthetic dataset and a real world dataset from the 2014-2016 Ebola outbreak, and compare the automatically created \SYSTEMNAME\ output to the manually generated state of the art in this community, the Nextstrain~\cite{hadfield_nextstrain:_2018} and Microreact~\cite{Argimon2016} applications, which also visualize Ebola data. 

\subsection{Comparison to ShowMe, Voyager, and Draco}
\label{subsec:eval-vis}
\tm{Although earlier recommendation systems such as ShowMe~\cite{mackinlay_show_2007}, Voyager~\cite{Wongsuphasawat2016}, and Draco~\cite{Moritz2019} are not designed to handle heterogeneous data types and are not optimized for the genEpi domain, we nevertheless compare to them as the closest previous work with respect to automatically created visual encodings.}   
%\rev{It is not entirely valid to compare \SYSTEMNAME\ to Showme, Voyager, and Draco  since, as we stated in the introduction, these systems have existing limitations that make them ill-suited to make recommendations for genEpi datasets.  Moreover, they are at a disadvantage to \SYSTEMNAME\, which is optimized for genEpi data. However, as these systems represent the state of the art we believe it is constructive to at least attempt to compare to them.}

\tm{Each of these systems is capable of making one or more recommendations for visual encodings provided a singular tabular dataset as input. Any genEpi data that is or could be transformed into a tabular dataset would thus be viable for these tools. It would be possible to pre-process multiple tabular dataset to combine them into a single tabular dataset in order to be usable by these systems. By comparison, \SYSTEMNAME\ is explicitly designed to handle multiple datasets and automatically creates a graph representation of these inputs, thus removing the need for the user to combine the data themselves. This graph representation of data is powerful not only for internal computation, but also for illustrating to the user the connections between their datasets in support of data recon and to help them verify data quality.}

\rev{Once datasets are loaded, all of these systems make fairly rapid recommendations through different mechanisms. 
ShowMe~\cite{mackinlay_show_2007}, Voyager~\cite{Wongsuphasawat2016}, and Draco~\cite{Moritz2019},
all use some notion of graphical efficacy to rank and prioritize visualization recommendations. 
ShowMe and Voyager
use a set of manually curated weights and a rule based system that is held fixed for all dataset. Draco uses a combination of user defined constraints and learned metrics of efficacy to prioritize visualization recommendations. To instantiate the Draco approach of learning metrics from perceptual experiments, one would need to conduct many additional experiments on visual encodings relevant to genomic epidemiology, such as phylogenetic trees, genomic maps, and even images, to make full use of such an inference engine. The user must also be able to programmatically specify constraints, which can be limiting to experts. In contrast, \SYSTEMNAME\ proposes a relevance metric that leverages a domain-specific visualization prevalence design space (VPDS) to capture and make use of visualization strategies employed by domain experts. One could image combining both domain-specific relevance and graphical efficacy metrics, in future work to further fine-tune visualization recommendations. All of these systems, \SYSTEMNAME\ included, presently require at least some human derived data or input to initiate the generation of visualization recommendations.}

\rev{Finally, we consider the types of visualizations that these systems produce. Considering the coverage of chart types, both Voyager~\cite{Wongsuphasawat2016} and Draco~\cite{Moritz2019} principally recommend common statistical charts such as scatter plots, bar charts, and histograms. ShowMe~\cite{mackinlay_show_2007} also recommends common statistical charts and its repertoire includes maps as well. None of these systems can make recommendations that include the trees, images, tables that are routinely visualized in the genEpi domain.} 

\tm{Considering the coordination across multiple charts, 
%these three systems  in isolation. 
ShowMe~\cite{mackinlay_show_2007} only generates a single visual encoding per Tableau worksheet. Although Tableau workbooks can have many worksheets, to combine visual encodings from individual worksheets into a dashboard requires manual action from the user, as does linking views to interactively filter data. 
To show users multiple different alternative designs, Voyager~\cite{Wongsuphasawat2016} produces more than one visual encoding and they may be linked through shared x or y axes or color. Users can pick and pin visual encodings they wish to review in greater detail later. However, these combinations are not deliberately created with the intent of coordination, but are instead an artifact of surfacing as many alternative views of the data as is reasonable. Draco~\cite{Moritz2019} can also programmatically produce multiple coordinated charts, but it is left up to user to facilitate the coordination process. The design of \SYSTEMNAME\ to explicitly produce coordinated chart combinations avoids the manual post-processing a user must undertake to combine and coordinate singleton charts with the other three systems.}

%\rev{Overall, these systems have many complimentary features that in concert address various challenges of automatically generating visualizations recommendations.}

\subsection{Comparison to Nextstrain and Microreact}
\label{sec:eval-genepi}
%and the Nextstrain and Microreact views in Appendix~\ref{app:comparison}. 
We now compare the results of \SYSTEMNAME\ to 
Nextstrain~\cite{hadfield_nextstrain:_2018} and Microreact~\cite{Argimon2016},
two human-curated visualization dashboards that have been developed by genEpi domain experts. 
All three of these systems feature a phylogenetic tree and a map. \SYSTEMNAME\ assigns high priorities to these over other chart types. Nextstrain  also shows a genomic map, while Microreact shows a timeline. While \SYSTEMNAME\ adaptively responds to input datasets, Nextstrain does not and relies on a specific analysis stack, while Microreact supports only two datatypes (tabular and tree) that can be loaded via the interface. Neither can show any other visual encodings. Nextstrain also shows a limited hand-curated set of field attributes regardless of the dataset, while Microreact allows user to select some fields to visualize. Nextstrain and Microreact can facilitate color aligned combinations, but these fields must be selected by the user or are hard coded by the developers. \SYSTEMNAME\ automatically chooses attributes to encode based upon a combination of data linkage and user specification, and these will vary according to the input data. However, some of \SYSTEMNAME's design choices are not as effective as Nextstrain and Microreact, especially the way that it chooses to encode temporal data. Also, data need not be loaded on external servers for \SYSTEMNAME\ to generate visualizations, in contrast to the other two systems.

\section{Evaluation with GenEpi Experts}\label{sec:eval-experts}
%\ac{\textbf{Biggest change from KDD is I brought back the CHI eval. I added a lot more numbers here because that was what one of the CHI reviewers really wanted. I personally liked the longer eval and there's room for it here.}}
We have conducted a qualitative study with genomic epidemiology experts to evaluate the usefulness, interpretability, and actionability of \SYSTEMNAME's visualization recommendations. We remind the reader that all study materials including anonymized study data are available in the Supplemental Materials and in our online repository: \url{https://github.com/amcrisan/GEVitRec}

\subsection{Study Procedures}
Study participants were walked through \SYSTEMNAME's visualization recommendation procedures through a chauffeured demonstration from the study administrator using the two datasets described in Section~\ref{sec:results}. The demonstration is documented in Supplemental Sec.~S2. 
%using publicly available data from the 2014-2016 Ebola Outbreak. The Ebola outbreak dataset has been described in the previous section. 
Chauffeured demonstrations showcasing both datasets were carried out using the RStudio IDE and run within an R Notebook environment. Participants were shown the top five views generated by \SYSTEMNAME\ for each dataset and were asked to provide a more detailed assessment and interpretation of the Ebola dataset views specifically. Participants could also interrupt the study administrator at any point in time to ask for clarification of either the algorithm's procedures or the data. Following the demonstration, participants were asked to complete an online questionnaire.

\subsection{Data Collection and Analysis}
We recruited ten genEpi domain experts to participate in our study. Participants were drawn from Canada, the United States, and the United Kingdom. One participant was a graduate student in genomic epidemiology and all others were professionals who either consulted or worked within regional or national public health agencies. Domain experts had varying degrees of exposure to genomic data, but all participants routinely analyzed heterogeneous data sources to conduct epidemiological investigations. The majority of experts identified as bioinformaticians and/or surveillance analysts. All experts had some exposure to the R language for statistical analysis and visualization (5 beginner, 4 intermediate, and 1 expert level of self-reported proficiency). Nearly all participants frequently developed data visualizations to understand and communicate their results. We collected data through an online questionnaire, administrator session notes, and audio recordings of the session. Results from all three sources were analyzed together. Anonymized data is available in the Supplemental Materials. When visualizing data, the majority of participants used R, Excel, Google Docs, or Tableau on a daily or monthly basis. Other tools that participants used were Microsoft Visio, Phandango, Power BI, Spotfire, and BioNumerics. 

%All participants had some exposure to R, with half identifying themselves to have intermediate and expert experience. We did not target R users in our recruitment; these demographics emphasize the importance of R in this community for analysis and visualization. Nearly all participants frequently developed data visualizations to understand and communicate their results. We collected data through an online questionnaire, administrator session notes, and audio recordings of the session. Results from all three sources were analyzed together. Anonymized data is available in the Supplemental Materials. %When visualizing data, the majority of participants used R, Excel, Google Docs, or Tableau on a daily or monthly basis. Other tools that participants used were Microsoft Visio, Phandango, Power BI, Spotfire, and BioNumerics. 

\subsection{Findings and Interpretation}
Supplemental Sec.~S3 contains all questionnaire results, Sec.~S4 the study administrator notes, and Sec.~S5 the transcripts of the session recordings. Here, we summarize participants' assessment of relevance and interpretability of our results.
%We summarize our study findings according to participant's assessment of GEViTRec's usability, the interpretability and relevance of its outputs, and their comparison of GEViTRec to other tools. We also summarize interesting findings participants surfaced, including the layout and summarization of results, how GEViTRec would be used for data reconnaissance, exploration, and targeted search, and finally how participants would use GEViTRec on their own data. 

\vspace{1mm}
\noindent\textbf{$\blacksquare$ Usability Assessment of GEViTRec}\\
Participants were asked to assess their perceived usability of \SYSTEMNAME\ from the chauffeured demonstration. 

\vspace{0.5mm}
\noindent\textit{\textbf{Overall, participants had a positive impression.}} Participants all strongly agreed or somewhat agreed they could quickly find a useful data visualization from the set of suggestions provided by the system and that it was much more useful for the system to automatically come up with visualizations than for them to come up with visualizations themselves. Participants indicated that GEViTRec visualized the data in ways they would not have thought of, but also indicated that GEViTRec did not show some of the visualizations they had expected. Their satisfaction with \SYSTEMNAME's results seemed to correlate with whether the participant felt they had a particular research question in mind: \textit{``I might have some very specific question that I want to answer. As opposed to when I am the researcher thinking about what is genomic epi of this disease. I might have much broader question which are address by this earlier image''}. Participants strongly agreed that GEViTRec produced relevant data visualizations that help them understand the data. 

\vspace{0.5mm}
\noindent\textit{\textbf{The entity graph is valuable for understanding data.}} Participants also strongly or somewhat agreed that it was understandable how the data connected together. In their discussion with the study administrator and their textual responses, they emphasized how the entity graph to be an especially useful way to understand their data: \textit{``the entity graph [...]is incredibly useful as well [...] it's really nice to see that visualization of how all of these different datasets can connect together''}. Participants indicated that they would like the entity graph to have more interactivity and would use it more centrally in their analyses. Their suggestions open the way to many potential future directions as GEViTRec continue to evolve.

\vspace{0.5mm}
\noindent\textit{\textbf{GEViTRec is fast and simple.}} Participants were impressed with the speed that data were linked and visualized. They also indicated that the limited amount of coding was manageable, but many felt that a dedicated user interface would have simplified it further and would increase GEViTRec's impact. Several participants also validated our claim that generating an R based tool was effective because of growing support and infrastructure in their organizations for R as an analysis tool. As such, GEViTRec could fit easily with their existing workflows. Finally, several participants asked about the data quality required to generate the data visualizations they were shown in through the demonstration. They were informed that GEViTRec acts as a data viewer and will visualize whatever quality of data is given. \tm{However, we suggested to participants this quality-agnostic approach of GEViTRec was a beneficial feature for data recon, especially when combined with the entity graph visualization, because it could help participants triage and wrangle their data within the R environment. Participants agreed that this GEVitRec use case usage would also fit within their workflows.} 

%\noindent\textit{\textbf{Provides a useful first glance of the data.}} Participants commented on the usefulness of GEViTRec to provide a very quick first glance of the data. We will elaborate on some of these finds a little more when discussing results pertaining to '', but generally, all participants were impressed with the ease and speed to generate visualizations and the ability to go from the data to 

\vspace{1mm}
\noindent\textbf{$\blacksquare$ Interpretability and Relevance of GEViTRec Output} \\
We assessed the relevance and interpretability of GEViTRec's results by asking participant to interpret its results \tm{on the Ebola dataset for the top ranked view}.

\vspace{0.5mm}
\noindent\textit{\textbf{Output and ordering is relevant and interpretable.}} All participants agreed that both the views generated by GEViTRec and the order they were presented in were useful and would help them with their current analysis. One participant stated that \textit{``these are all extremely useful visualizations. And having essentially a menu to choose from is fantastic'}'. The interpretability of the views was further by validated by  participants' ability to dive into the interpretation of the Ebola outbreak data and begin answering questions of \textit{``when, how much, and where''} the outbreak occurred. Some participants commented on the redundancy of information between the chart types saying that the views could more effectively communicate the diversity of data types and attributes. However, some participants stated that the redundancy of information encoding in the charts was effective because it allowed them to \textit{``tease out different aspects of it [data] that you might be interested in''}.

\vspace{0.5mm}
\noindent\textit{\textbf{Addressing issues of data scale.}} The majority of participants indicated the Ebola data was more difficult to interpret than the synthetic data, a surprise to us due to the well-publicized nature of that outbreak. 
%This was a surprising, because the Ebola outbreak was so publicly advertised and known we though that participants would have a very easy time interpreting. 
However, participants indicated the amount of data and variety of chart types together was overwhelming: \textit{``It's not as easy to process as the sample [synthetic] data was, which you could take one look at it and say, I totally understand everything that's going on. This one would take me a few more minutes of just sitting and thinking and needing to digest a little bit.''} Data aggregation, as Draco~\cite{Moritz2019} or Voyager~\cite{Wongsuphasawat2017} does, is one potential solution, and GEViTRec does so for certain data and chart types  such as bar charts. However, aggregating procedures for more complex data types, like phylogenetic trees, are not trivial; domain expertise is currently required. Below, we discuss how layout, effective text summaries, and improved chart coordination may help address these issues.

\vspace{1mm}
\noindent\textbf{$\blacksquare$ Comparison to Nextstrain and Microreact}\\ 
Participants were shown the Ebola outbreak dataset on the Nextstrain~\cite{hadfield_nextstrain:_2018} and Microreact~\cite{Argimon2016} applications and were asked to discuss their perceived differences between human generated (Nextstrain, Microreact) and machine generated (GEViTRec) views.

Participants felt that the Ebola data was overwhelming, regardless of the data visualization tool, and that they needed more help understanding how to approach and consume the information. However, participants appreciated that Nextstrain and Microreact had interactive components that allowed them zoom in and filter the data. Future work could allow this ability for GEViTRec if it were embedded with an R Shiny interface. Several participants liked Nextstrain's animation of disease transmission. They also felt that compared to GEViTRec, the human curated visualizations were more aesthetically pleasing and more carefully coordinated. However, they felt that GEViTRec contained information that the human generated views did not. Participants also felt that both Nextstrain and Microreact were suited for very specific tasks, and not helpful if you wanted to see a different view of the data or had a different research questions :\textit{``phylogeography is not very interesting to what I am doing, so instantly one-third of that [...]  visualization is not that useful''}. Participants indicated that it was important for data visualization tools to adapt to different datasets as GEViTRec does.
%Finally, they felt that data visualization tools should be able to adapt and show different views dependent upon the input data, and said they appreciated GEViTRec's adaptability to different data.

\vspace{1mm}
\noindent\textbf{$\blacksquare$ Layout and Summarizing of Results}\\
%We primarily set out to assess GEViTRec's algorithmic results and to verify that it is capable of producing interpretable and relevant visualizations for experts. However, experts raised a number of points of how the layout and representation of the data visualizations would help them better interpret the data. 
Participants liked that different chart types were linked by spatial or color alignment, or by having common attribute fields on the axes. However, they felt that data could be presented more effectively to help them consume the information, especially in the Ebola data example where there was so much information. Several participants said it would be helpful to have text summaries that describe the data and the type of visualization, a finding echoing a prior study on generating a report design for genomic epidemiology data~\cite{peerJ2018}. Participants also thought it could be helpful to lay out data in a `scrollytelling' vertical layout, and one suggested that information could be further prioritized to show the simplest data first and more complex data (and their attendant visualizations) later. Participants also indicated that different charts should be sized according to their data density and relative importance. For example, several participants felt that the phylogenetic tree should be large in the Ebola example, both because there was so much data and because it was important, whereas the bar chart should have less screen real estate because, even though the information was important, the `data-ink' ratio was not justified. 

%\vspace{0.25mm}
%\noindent\textit{\textbf{Using color more judiciously.}} Participants appreciated that information was linked across different chart types, via linking attribute fields, using color. However, they also indicated that it could be confusing when more than one attribute is used to encode color. Moreover, they indicated we need to have a more careful selection of colour palettes so that they were not conflicting. Participants indicated that when they considered the information for a longer period of time, that it was possible to interpret all of the different chart links facilitate through colour encodings, but at first glace our default choices were ineffective.  

\vspace{1mm}
\noindent\textbf{$\blacksquare$ Value for Data Reconnaissance}\\
\SYSTEMNAME\ was designed to provide users with a quick first glance of the data so that they can assess its value or the need to pursue additional datasets. Participants agreed that GEViTRec fulfilled this intended data recon purpose: \textit{``This tool is extremely useful for data exploration, particularly where there are incomplete, or highly varied types of data that must be integrated and displayed. Very useful for public health surveillance and initial review of data.''} Participants also repeatedly stated they saw \SYSTEMNAME\ as a useful hypothesis generation mechanism, helping them to see multiple views of their data that they had not considered. Even more interestingly, many participants differentiated between \SYSTEMNAME\ functionality and the much more targeted search of generating a specific data visualizations for a specific question; essentially \SYSTEMNAME\ was seen as less effective for hypothesis verification but useful for hypothesis generation. Hypothesis generation more closely aligns with the goals of data recon than hypothesis verification, however, it would be fruitful to explore this particular tension in follow-on work. Ultimately, participants found the value in both being able to generate a specific visualization and to see alternative visualizations: \textit{`` when you're discovering things, then maybe it is good to have things that are generated without human input.''}. 

\vspace{1mm}
\noindent\textbf{$\blacksquare$ Further customization to participant background}\\
All participants wanted to use \SYSTEMNAME\ on their own data and analyses. When queried about more specific datasets they would apply \SYSTEMNAME\ to, all participants identified data with a large number of samples, sourced from different data types, and with large numbers of diverse attribute fields. However, several participants wanted further optimization of the relevance ranking to include participant's background. Interestingly, participants also stated that they still wanted to see the multiple views of the data to preserve the ``hypothesis generation'' and ``exploratory'' abilities that perceived \SYSTEMNAME\ afforded: \textit{`` relevance is very specific for that [genEpi] person. [...] having that [research question] as a criteria [...] would be very helpful to target your visualizations. [...] there's the counter point that it's sometimes helpful to see things that you are not asking the question for. So that the top ranked view was more tailored to them, but that lower ranked views needed not necessarily be.''}

\section{Discussion}
Heterogeneous and multidimensional data are already the norm in many domains and stakeholders are increasingly expected to use these complex data to derive informed actions~\cite{gray:2009}. In our own collaborations we have seen stakeholders struggling to understand landscapes of heterogeneous data, and we believe that these challenges are not limited to genEpi, but extend more broadly to other data science applications. \rev{We developed the terminology of data recon and a conceptual framework of how to tackle it in prior work~\cite{DRTW_crisan_2019} to capture the challenges of these experts and to delineate their unmet needs. Our prior conceptual framework identified a four-phase iterative cycle for data recon (acquire, view, assess, pursue) and emphasized the importance of experts being able to rapidly see visual encodings that provide an overview of their data. While we identified visualization recommender systems as a viable solution to the challenges of data reconnaissance, we also identified the limitations of previous systems that lower their utility to experts in genEpi and other domains with heterogeneous data landscapes.}
%We were motivated by these challenges to develop a novel techniques for visual data reconnassiance (\TECHNIQUE), which builds upon prior work of visualization recommender systems. In addition to this technique, we also contribute \SYSTEMNAME\ as  proof-of-concept of how our technique compares to existing systems, including visualization created by genEpi experts, and how experts themselves respond to the suggestions to the system.  
The recommender algorithm that we present here is a significant advance over the previous state of the art that addresses the challenges of data recon in complex and diverse data landscapes. \tm{We demonstrate that the \SYSTEMNAME proof-of-concept implementation provides rapid overviews across a wide variety of data types and produces actionable insights that experts can interpret and integrate into their existing workflows. Our evaluation provided evidence that our approach of using coordinated combinations of static charts was a viable approach to data recon, supporting very fast overview comprehension without requiring the time investment of interactivity.   }

\subsection{Domains Beyond GenEpi} 

\tm{The \SYSTEMNAME\ algorithm has many domain-agnostic elements, but also incorporates information from a domain-specific visualization prevalence design space (VPDS). While our specific implementation was in the genEpi domain, we intended this approach to be transferrable to other domains as well. The future work needed to fully validate our claim of domain independence would be to validate and deploy an implementation in another domain. However, one current bottleneck in doing so would be the manual effort currently required to generate a VPDS. While the GEViT method for doing so requires a mix of automatic computation and human effort~\cite{gevit2018}, a robust method that is fully automatic would address that problem.} 

\tm{
We do not solve the technical problem of automatically generating a VPDS in this work. 
Instead, what we demonstrate here is how a design space can be used to inject domain expertise into recommendation systems. 
We hope that our work makes headway on this  ``chicken and egg'' problem, motivating further work on automating the construction of a VPDS by showing the practical benefits that can be obtained from having one available.} 
%While we focus here on the genEpi domain, our technical contributions can be deployed to other contexts. However, this transfer of our results requires additional automation. One bottleneck is the effort involved in generating domain prevalence design spaces. While a previously proposed method for doing so requires a mix of automatic computation and human effort~\cite{gevit2018}, a robust method that is fully automatic would address that problem. 
\tm{Moreover, our current approach also requires further manual analysis of VPDS results, such as the chart combination viability matrix, which would also benefit from automation.}   
A more prosaic problem to extending this work to other domains, where there is no clear route for automation, is the amount of tedious and time-consuming work required to integrate domain-specific software packages that visually encode new data types into a software system. %Alleviating these bottlenecks and developing instantiations in other domains will be an exciting area of future work. % For example, the link between the underlying data and resulting visualization. 
\tm{Our current \SYSTEMNAME\ implementation is a proof-of-concept that such an end-to-end system, where the input of multiple heterogeneous data sources results in the automatically generated output of domain-relevant coordinated combinations of charts, is viable and that further automation would be a worthwhile investment to extend it to other domains.} 

\subsection{The Relevance of Prevalence}
%\ac{Added this fun section back from the CHI paper}.
Our algorithm is built to prioritize chart types according to a relevance metric that is built on domain-specific prevalence, namely examining the commonly used visualization strategies of domain experts, in contrast to the many existing recommender systems use graphical perceptual effectiveness to rank visual encodings. 
We point out the concern that relying on perceptual effectiveness as the sole ranking mechanism may not be sustainable at scale because of the large number of studies that would be necessary even to assess single charts, in light of the full range of visual encodings possible for a heterogeneous array of data types. Moreover, it is even more challenging to fully assess the perceptual implications of combinations of charts, because these combinations introduce many perceptual questions that are challenging to isolate in an experiment. Using information from a domain-specific visualization prevalence design space does not preclude incorporating efficacy judgements. It would be fruitful future work to examine the trade-offs between perceptual effectiveness and domain prevalence, and how to combine them.  If perceptual experiments provide adequate coverage for any specific visualization prevalence design space, then it would be possible to penalize relevant visualizations that are not perceptually effective. 

One potential objection to defining relevance according to domain-specific prevalence is that domain experts are not visualization experts and thus may visualize their data inappropriately, without a full awareness of the breadth of possible visualization designs and the trade-offs between them. Although we agree that individual domain experts may not be fully aware of how to visualize their data, we have observed that the collective strategies of a large group of experts can reveal a complex combinatorial design space. When we use this domain-specific prevalence information in a recommender systems, we can generate visual encodings that individual experts find intriguing and informative although they believe that they would not have generated them manually. As a preliminary measure, the prior GEViT study tagged some visualizations as being ``good'' or ``missed opportunity'' based upon subjective expert judgement~\cite{gevit2018}. These measures were not incorporated in our relevance metric, but this information could be generated at a larger scale and incorporated into future iterations of \SYSTEMNAME. 

Our approach represents only one of many possibilities, and future visualization recommender systems may contain several different ranking metrics that may be tuned and optimized through some combination of machine learning and user preference.

\vspace{-1.5mm}
\subsection{Future Work}
In the process of developing and implementing \SYSTEMNAME\ we identified several areas of important future work. In the very short term, we can continue to improve how \SYSTEMNAME\ lays out its results and further tailors visualization output to an expert's context. The addition of narrative elements, such as text, or staged displays of the data, could help to alleviate some challenges interpreting visualizations containing a lot of data.  In the longer term, there remain several interesting and challenging issues in generating coordinated combinations, including integrating more complex types of combinations and incorporating interactivity. Finally, our research has interesting potential to inform human interactions with machine learning or artificial intelligence systems~\cite{heer:2019}. Many of our experts already raised relevant discussion points of how they interact with the results of our system. It would also be fruitful to consider mixed-initiative systems where the initial visualizations are automatically generated and then interactively modified by the user, with the underlying algorithm learning from user refinements. Such a system could significantly reduce the time it takes to complete data exploration and quality assessment tasks in analytic workflows, and would help stakeholders rapidly identify potentially actionable insights in their complex data landscapes.

\section{Conclusion}
We have presented a novel algorithm for data reconnaissance through visualization recommendation, \SYSTEMNAME, that automatically generates coordinated combinations of charts from multiple heterogeneous datasets. Our approach includes many domain-agnostic components, but also incorporates information about the prevalence of visualization usage within a specific target domain, in our case genomic epidemiology. We created a proof-of-concept implementation to demonstrate its capabilities using both real and synthetic data, and conducted a thorough qualitative evaluation with ten domain experts to validate our claims that our method quickly produces relevant and interpretable views of data. 

%Through our studies we also identified important considerations for the generation of visualization recommendations that can transfer to other domains struggling with similarly complex data landscapes. 

%Our approach represented one of many possible strategies for generating a visualization design space. Other examples include SetVis~\cite{setvis2014}, TreeVis~\cite{schulz_treevis.net:_2011}, ManyEyes~\cite{manyeyes}, the Tableau Public analysis~\cite{Morton:2014}, and VizNet~\cite{Hu:2019:VTL:3290605.3300892}. However, compared to this prior work our representative sampling strategy allowed us to enumerate these different visualization strategies and obtain a quantitative sense of their relevance and importance to genomic epidemiology; information we use in \SYSTEMNAME. 

% or


%\appendices

%\section{}

\ifCLASSOPTIONcompsoc
  \section*{Acknowledgments}
\else
  \section*{Acknowledgment}
\fi

We thank the members of the UBC InfoVis group for their thoughtful comments and feedback. We also acknowledge and thank our study participants for their time and insights.

\ifCLASSOPTIONcaptionsoff
  \newpage
\fi



\bibliographystyle{IEEEtran}
\bibliography{sample.bib}


%%Biography text here.
%\end{IEEEbiography}

%\ac{I really don't want to add pictures, that always seemed silly to me}.
\begin{IEEEbiography}[{\includegraphics[width=1in,height=1in,clip,keepaspectratio]{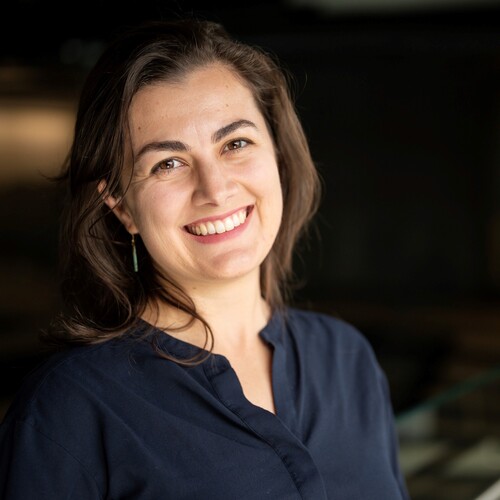}}]{Anamaria Crisan}
is a Research Scientist at Tableau. She completed her PhD in Computer Science in 2019 at the University of British Columbia. Her research explores visualization through the data analysis life cycle with an emphasis on biomedical applications. Her work on this project was supported by a CIHR Vanier Award. 
\end{IEEEbiography}

\begin{IEEEbiography}[{\includegraphics[width=1in,height=1in,clip,keepaspectratio]{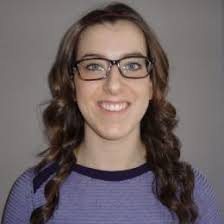}}]{Shannah E. Fisher}
is an undergraduate student in Computer Science at the University of British Columbia.
\end{IEEEbiography}

%\newpage

\begin{IEEEbiography}[{\includegraphics[width=1.02in,height=1in,clip,keepaspectratio]{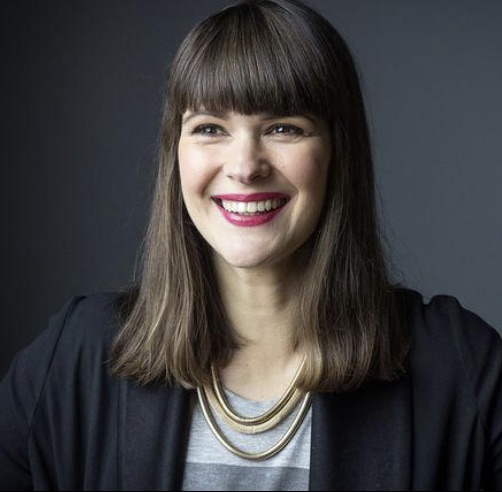}}]{Jennifer L. Gardy}
is the deputy director, Surveillance, Data, and Epidemiology for the Malaria team at  the Bill \& Melinda Gates Foundation. Before that, she spent ten years at the BC Centre for Disease Control and the University of British Columbia's School of Population and Public Health, where she held the Canada Research Chair in Public Health Genomics. Her research focused on the use of genomics as a tool to understand pathogen transmission.
\end{IEEEbiography}

\begin{IEEEbiography}[{\includegraphics[width=1in,height=1.25in,clip,keepaspectratio]{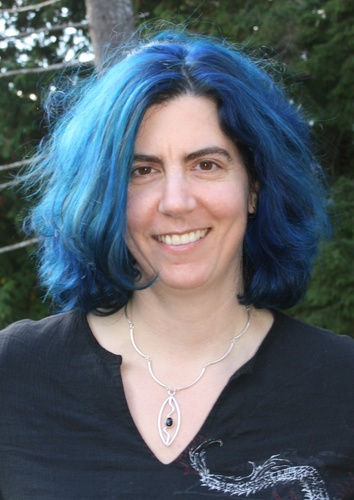}}]{Tamara Munzner}
is a professor at the University of British Columbia, and holds a PhD from Stanford from 2000. She has co-chaired InfoVis and EuroVis, her book Visualization Analysis and Design appeared in 2014, and she received the IEEE VGTC Visualization Technical Achievement Award in 2015. She has worked on visualization projects in a broad range of application domains from genomics to journalism.
\end{IEEEbiography}


\vfill

% Can\rev{We observed in prior studies~\cite{peerJ2018,DRTW_crisan_2019} that domain experts in gen be used to pull up biographies so that the bottom of the last one

\end{document}